\pdfoutput=1
\documentclass[preprint]{rmaa}
\usepackage[utf8]{inputenc}
\usepackage{url}
\usepackage{graphicx}
\usepackage{color}
\usepackage{placeins}
\usepackage{float}
\usepackage{tabularx,colortbl}
\usepackage{ifthen}
\usepackage[table,xcdraw]{xcolor}

\usepackage{amsmath}
\usepackage[english]{babel}
\usepackage{verbatim}
\usepackage{multirow, array}
\usepackage{tabulary}
\usepackage{paralist}
\title{Blade Antenna-SDR System Prototype for the CANTAR Global 21-cm Experiment: Simulations, Measurements, and In-Situ Results} 

\author{
  Felipe P. Mosquera\altaffilmark{1,4}, 
  Julian Rodriguez-Ferreira\altaffilmark{1},
    Efrén Acevedo\altaffilmark{1},\\
  Oscar Restrepo\altaffilmark{2,4},
  David González\altaffilmark{1},
  \& Germán Chaparro\altaffilmark{3,4}}

\altaffiltext{1}{Escuela de Ingenierías Eléctrica, Electrónica y Telecomunicaciones. Universidad Industrial de Santander,Universidad Industrial de Santander, Bucaramanga, Colombia.}

\altaffiltext{2}{Universidad ECCI, Bogot\'{a}, Colombia. }

\altaffiltext{3}{Grupo de Física y Astrofísica Computacional, FACOM, Universidad de Antioquia, Medell\'{i}n, Colombia.}

\altaffiltext{4}{Fundación para el Desarrollo de la Radioastronomía y Tecnologías Aplicadas - FUDARTA, Bogotá, Colombia.}

\shortauthor{Mosquera et al.}
\shorttitle{Blade Antenna-SDR System Prototype for 21-cm Cosmology}

\fulladdresses{
\item Efren Acevedo, David Gonzalez, Felipe P. Mosquera, Julian Rodriguez-Ferreira: Universidad Industrial de Santander, 680002, Carrera 27 Calle 9, Bucaramanga, Santander, Colombia (efrenacevedoc@gmail.com, david2248098@correo.uis.edu.co, felipe.mosquera@fudarta.org, jgrodrif@uis.edu.co).

\item German Chaparro: Universidad de Antioquia, Calle 67 N. 53-108, 050010, Medellin, Antioquia, Colombia (german.chaparro@udea.edu.co).

\item Oscar Restrepo: Universidad ECCI,  Carrera 19 N. 49 - 20, 110221, Bogot\'{a}, Cundinamarca, Colombia (orestrepog@ecci.edu.co).}

\listofauthors{ F.\~P. Mosquera, J. Rodriguez-Ferreira,  E. Acevedo, O. Restrepo, D. Gonzalez, G. Chaparro}
\indexauthor{Mosquera et al.}

\abstract{We present the design and initial testing of a low-frequency radio telescope prototype developed for the CANTAR (Colombian Antarctic Telescope for 21-cm Absorption during Reionization) experiment. Operating from 100 to 200 MHz, the system integrates a blade dipole antenna inspired by the EDGES high-band design with a software-defined radio (SDR) receiver. We report simulations of antenna impedance and beam chromaticity, along with SDR performance tests (Limenet Mini, Ettus E310, USRP2920). A dual-stage low-noise amplifier reduces system temperature, enabling foreground-sensitive observations. Radiometric estimates suggest sub-mK sensitivity is achievable with 1000 h of integration. This prototype forms part of Colombia’s emerging infrastructure for 21-cm cosmology, with deployments planned in low-RFI sites in the Colombian Andes and Antarctica.}

\resumen{Presentamos el diseño y pruebas de un prototipo de radiotelescopio de baja frecuencia para el experimento CANTAR (Colombian Antarctic Telescope for 21-cm Absorption during Reionization). El sistema opera entre 100--200 MHz, y combina una antena blade dipolo, inspirada en el diseño de EDGES-HB, con un receptor basado en radio definido por software (SDR). Reportamos simulaciones de impedancia y cromaticidad del haz de la antena, junto con pruebas de desempeño de varios SDRs (Limenet Mini, Ettus E310, USRP2920). Un amplificador de bajo ruido de dos etapas reduce la temperatura del sistema, permitiendo observaciones sensibles a la emisión de primer plano. Estimaciones radiométricas indican que es posible alcanzar sensibilidad sub-mK con 1000 h de integración. Este prototipo es parte de la infraestructura emergente en Colombia para cosmología de 21 cm, con campañas planeadas en sitios de bajo RFI en los Andes Colombianos y la Antártida.}

\addkeyword{cosmology: observations}
\addkeyword{dark ages, reionization, first stars}
\addkeyword{instrumentation: detectors}
\addkeyword{instrumentation: miscellaneous}
\addkeyword{radio continuum: general}
\addkeyword{site testing}
\begin{document}
% Typeset article header
\maketitle

\section{Introduction}
\label{sec:intro}

\noindent The redshifted 21-cm line of neutral hydrogen provides a unique observational window into the thermal and ionization history of the early universe. In particular, measurements of the global (sky-averaged) 21-cm signal can probe the Cosmic Dawn and the Epoch of Reionization (EoR)  \citep{barkana, Furlanetto_2006, 2016MNRAS.455.3890M}. During this period, the first luminous sources coupled the hydrogen spin temperature to the kinetic temperature of the gas via the Wouthuysen-Field effect, producing an absorption signature in the 21-cm line against the cosmic microwave background (CMB) \citep{costa}. The frequency range corresponding to this evolution lies between approximately 40 and 200 MHz for redshifts $6\leq z \leq 30$ \citep{article6}, with an amplitude of $\sim$100–200 mK and width of $\sim$10 MHz, depending on the astrophysical heating history. \\

This signal is embedded in a sky dominated by bright galactic synchrotron emission and extragalactic foregrounds that are several orders of magnitude stronger ($\sim10^3-10^4$ K). Its detection demands tight control over instrumental systematics, including beam chromaticity, impedance mismatches, and thermal gain variations \citep{bowman2009,bernardi, Monsalve_2017, Cheng, restrepo}. The controversial EDGES detection of an absorption feature at 78 MHz \citep{2bc8363417fb44159194d3775d6e51bd}, with an amplitude exceeding theoretical expectations, has renewed interest in independent validation and motivated a new generation of experiments \citep{mist2023}.\\

Most global 21-cm efforts use single-element radiometers with fixed analog backends. These include EDGES, SARAS \citep{Singh_2018}, PRIZM \citep{prizm}, REACH \citep{reach}, BIGHORNS \citep{Sokolowski_2015}. Though differing in calibration approach, antenna type, and beam chromaticity control, they share the goal of suppressing spectral structure in the instrument response. Interferometric arrays such as HERA \citep{DeBoer_2017}, LEDA \citep{10.1093}, and LOFAR \citep{6051249} pursue complementary spatially resolved 21-cm measurements.\\

In this context, we present the design and initial testing of a modular, low-frequency radio telescope prototype developed at the Universidad Industrial de Santander (UIS). The system combines a blade dipole antenna, modeled after the EDGES high-band design, with a reconfigurable software-defined radio (SDR)–based receiver chain.\\

Unlike purpose-built analog backends, SDRs provide flexibility and rapid prototyping capabilities in radio science applications, but raise concerns about spectral fidelity due to ADC resolution, phase noise, clock instability, and internal filtering \citep{article1,sdrragoo,7303211}. One aim of this work is to quantify these limitations and assess whether SDRs can serve as viable alternatives for radiometer-based 21-cm cosmology when combined with analog filtering and appropriate calibration strategies.\\

We also present electromagnetic simulations of the antenna, including $S_{11}$ and beam chromaticity analysis, as well as in situ measurements of gain, sensitivity, and linearity for different SDR devices. We evaluate the dynamic range and noise temperature of the system under different preamplifier configurations and explore its response to controlled input signals. Although this prototype does not yet attempt sky signal extraction, it establishes a path toward future foreground characterization and calibration campaigns. \\

This prototype is the first instrument developed under the CANTAR initiative (Colombian Antarctic Telescope for 21-cm Absorption during Reionization), a long-term effort to design, test, and deploy low-frequency radiometers for global signal studies in low-RFI environments. CANTAR combines electromagnetic modeling and optimization with the development of novel analog and digital systems. It also features a robust Bayesian signal extraction and validation pipeline, along with comprehensive site testing. Future deployments are planned for high-altitude locations in the Colombian Andes, as well as in Antarctica as part of the Programa Radioastronómico Antártico Colombiano (PRAC).

This paper is structured as follows. Section 2 describes the system architecture, including the antenna design, SDR receiver configuration, and measurement setup. Section 3 presents the results from electromagnetic simulations, SDR performance tests, and in-situ measurements. In Section 4, we discuss the comparison between simulations and measurements, and assess the prototype's performance relative to other 21-cm experiments. Section 5 outlines future work and Section 6 summarizes our main findings and conclusions.

%This project is supported by ongoing related initiatives that seek to consolidate the UIS as a national reference in the study of radio astronomy. This system will merge with a radio astronomy station of UIS framed in the project \textit{"Desarrollo de un arreglo interferom\'{e}trico de Radio Telescopios para establecer una estaci\'{o}n de Radio Astronom\'{i}a de la UIS en el P\'{a}ramo de Berl\'{i}n (Santander)"}. Also, it will be used for the project \textit {"Caracterizaci\'{o}n de emisiones electromagn\'{e}ticas de interferencia de radiofrecuencias y pruebas de un radio telescopio (100mhz) como insumos para la validaci\'{o}n del sitio de montaje de una base radio astron\'{o}mica en la pen\'{i}nsula ant\'{a}rtica (fase cero dentro del Programa Radio Astron\'{o}mico Ant\'{a}rtico  Colombiano (PRAC)"}. Both initiatives will contribute to the country's scientific and technological development.

\section{Methods}

\subsection{System Requirements}

The design of the prototype system presented here is guided by constraints common to global 21-cm signal experiments like suppressing spectral structure in the system response and achieving wideband impedance matching. While the antenna does not require high spatial resolution, it must provide maximum gain at zenith and maintain a return loss better than 10~dB, gain between 10-30~dB, and a noise figure below 4~dB \citep{restrepo}. Table~\ref{tab:tabla} summarizes the relevant characteristics of comparable global signal experiments.\\

At frequencies below 100~MHz, ionospheric refraction and absorption add further spectral complexity. These effects are not directly addressed in this study, but must be accounted for in future long-duration deployments (Mora et al., in prep).

\subsection{Role of SDR Technology}

Software-defined radios (SDRs) provide a reconfigurable, low-cost backend with tunable sampling rates, bandwidths, and gain. Despite their widespread use in RFI monitoring and radiometry \citep{7303211, osti_1074425}, SDRs remain underused in global 21-cm experiments due to their limited dynamic range, internal filtering stages, and susceptibility to quantization and clock drift artifacts.\\

In this work, we evaluate two commercially available SDRs—the LimeNet Mini and Ettus E310—under conditions relevant for low-frequency radiometry. Both support open-source tools (GNU Radio, UHD) and are capable of tuning across the 100–200~MHz range. We assess  the suitability of each device  through controlled signal injection, focusing on gain linearity, spectral flatness, and noise floor stability.

\subsection{Radiotelescope System Description}

Our prototype follows a modular design that combines a broadband blade dipole antenna, a flexible analog front end, and an SDR-based digital receiver. While the antenna and ground plane geometry are inspired by the EDGES high-band system, our implementation introduces SDR-based signal processing and open-source data acquisition, allowing rapid reconfiguration and field deployment.\\

\begin{figure}[h]
    \centering
    \includegraphics[scale=0.5]{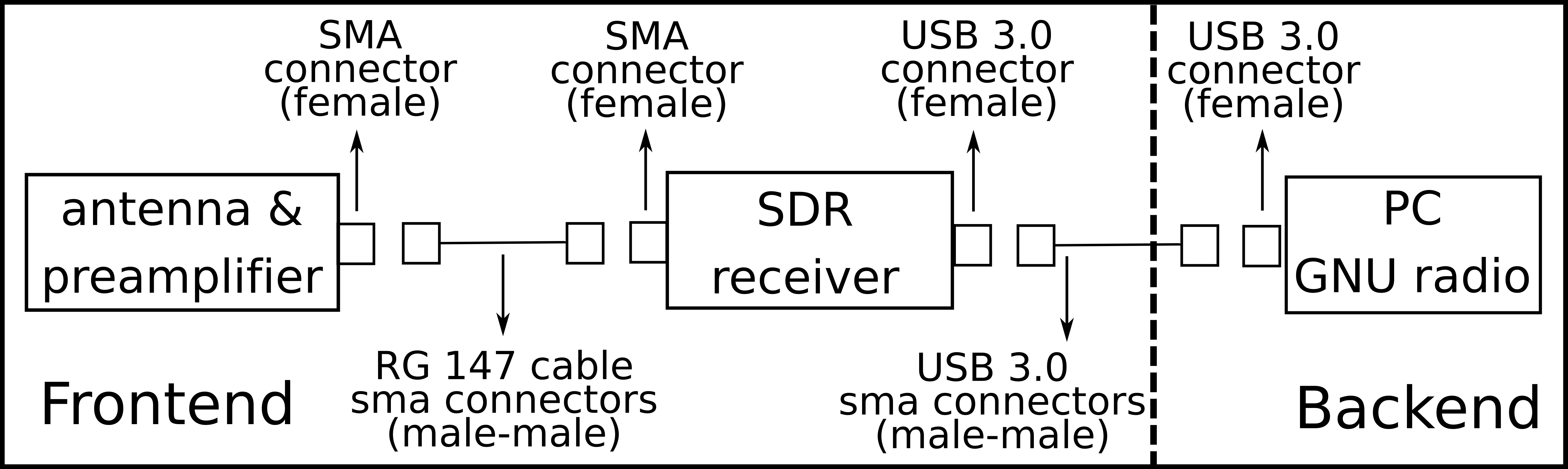}
    \caption{Subsystems of the radiotelescope to be validated.} 
    \label{fig:scheme}
\end{figure}

The system comprises three main subsystems: (1) a blade antenna mounted over a wire-mesh ground plane; (2) an analog front end including a balun and cascaded LNAs; and (3) a digital backend based on commercial SDRs, with acquisition scripts tailored for the platform in use. A system diagram is shown in Figure~\ref{fig:scheme}.

\begin{table}[h]
\centering
\caption{Comparison between 21-cm cosmology experiments}
\label{tab:tabla}
\begin{tabular}{ccccc}
\hline
\textbf{System} & \textbf{\begin{tabular}[c]{@{}c@{}}Frequency \\ range (MHz)\end{tabular}} & \textbf{\begin{tabular}[c]{@{}c@{}}Antenna \\ type\end{tabular}}         & \textbf{\begin{tabular}[c]{@{}c@{}}Use of \\ balun\end{tabular}} & \textbf{Calibration scheme}                                                                                                             \\ \hline

SARAS2          & 40-200                                                                    & spheric monopole                                                         & no                                                               & coupled noise source                                                                                                                    \\
 
EDGES           & \begin{tabular}[c]{@{}c@{}}100-200 \\ 50-100\end{tabular}                 & blade                                                                    & yes                                                              & \begin{tabular}[c]{@{}c@{}}antenna and noise source \\ commutation\end{tabular}                                                         \\

BIGHORNS       & 70-200                                                                    & cone log-spiral                                                          & yes                                                              & \begin{tabular}[c]{@{}c@{}}antenna and reference source \\ commutation\end{tabular}                                                     \\
 
SCI-HI          & 40-130                                                                    & hibiscus                                                                 & yes                                                              & \begin{tabular}[c]{@{}c@{}}antenna, 50 and 100 Ohms \\ loads and short terminal \\ commutation\end{tabular}                             \\

LEDA            & 40-85                                                                     & \begin{tabular}[c]{@{}c@{}}double \\ polarization \\ dipole\end{tabular} & yes                                                              & \begin{tabular}[c]{@{}c@{}}antenna and noise \\ source commutation combined\\  with cross-correlation \\ of other antennas\end{tabular} \\

MIST          & 40-125                                                                    & blade dipole                                                                 & yes                                                              & \begin{tabular}[c]{@{}c@{}}reference source calibration \\ and soil calibration\end{tabular}                             \\

\end{tabular}

\end{table}

\subsubsection{Antenna and Ground Plane}
\begin{figure}[h]
    \centering
    \includegraphics[scale=0.5]{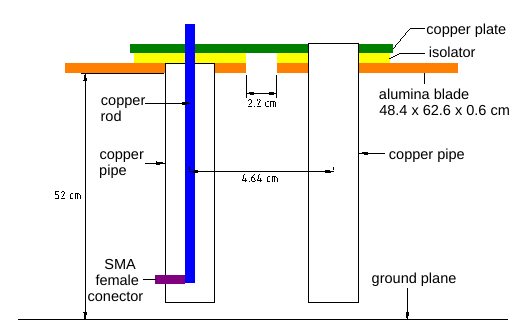}
    \includegraphics[scale=0.2]{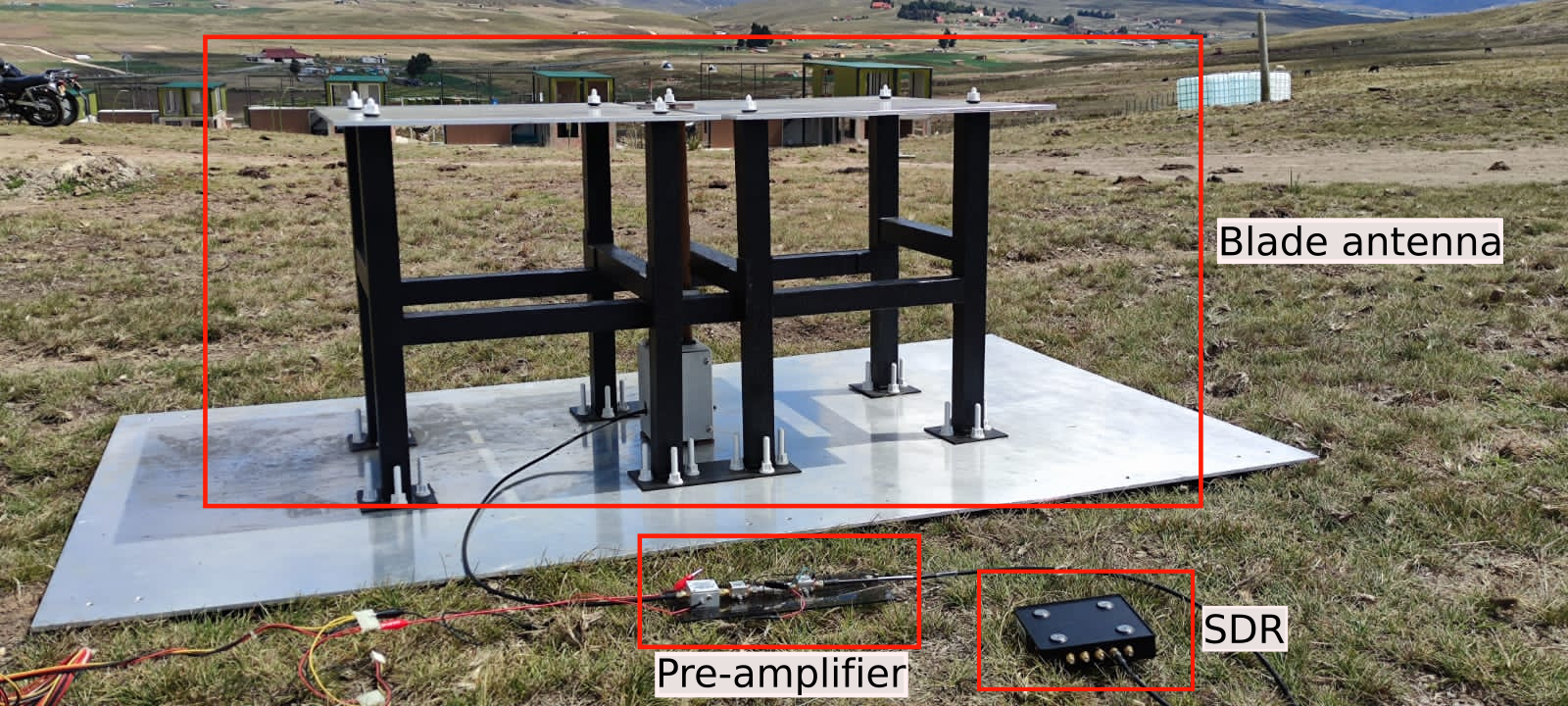} 
%    \includegraphics[scale=0.3]{im/tuner.jpg}\\
%    \raggedright
%    \figurenote{\textit{Note.} Schematic of the antenna (top), image of the blade antenna (middle)}
    %and closeup of the balun tuner (bottom).}
    \caption{100-200 MHz blade antenna. Schematic of the antenna (left), blade antenna in-situ at Páramo de Berlín (right).} 
    \label{fig:ant_1}
\end{figure}

The antenna is a single-polarization blade dipole, constructed from two aluminum plates mounted above a reflective ground plane. Its geometry is based on the high-band design used in the EDGES experiment, optimized for the 100–200 MHz frequency range following the Particle Swarm Optimization (PSO) scheme in \citep{restrepo}. The dipole elements measure 62.5$\times$48.1 cm and are positioned 52 cm above the ground, supported by a polystyrene structure chosen for its low dielectric constant and ease of fabrication (see Figure \ref{fig:ant_1}).\\

The ground plane consists of a central 1.3$\times$1 m aluminum sheet, extended using wire mesh panels to a final area of 5$\times$5 m. While initial simulations used only the solid plate, all measurements were performed with the full extended plane, minimizing edge diffraction and reducing ground reflection systematics. Although this ground configuration is relatively modest compared to those used in experiments targeting precision spectral calibration, it represents a practical compromise for a prototyping phase.\\

A Roberts (or Collins) balun \citep{inbook1100} connects the balanced dipole to the unbalanced coaxial transmission line. This balun consists of two copper tubes and a sliding tuning plate, which allows for mechanical adjustment of the electrical length and impedance match. By varying the tuner height, the resonant frequency and bandwidth of the antenna can be shifted. A metal enclosure was added to suppress common-mode currents along the feed structure. The output port is a standard SMA connector located at the balun base.

%\begin{figure}[h]
%    \centering
    %\caption{Schematic of the Roberts Balun.} 
%    \includegraphics[scale=0.35]{Images/bal_graph.png} 
%    \caption{Schematic of the Roberts balun. $A$ represents a 1/4 wave transformer (transmission line), $B$ a shorted balanced line, $C$ an open stub line (open-ended line), and $D$ the tuner. $Za$, which is the antenna impedance, is in parallel with $B$. The unbalanced port is introduced through $A$ and connected in series to the antenna by $C$. The lengths of $A$ and $B$ can be reduced by $D$. Adapted from \citep{inbook1100}.} 
%    \label{fig:tuner2}
%    \raggedright
    %\figurenote{\textit{Note.} Adapted from \cite{inbook1100}.}
%\end{figure}
%The impedance of B is defined by $ZB=R_b$ (1+$\gamma$)/(1-$\gamma$), where $\gamma=-e^{i\phi_b}$, $R_b$ is the line impedance, $\phi_b=-4\pi l \epsilon ^\frac{1}{2} f/c$  (being l the length, f the center frequency, c the velocity of light and $\epsilon$ the dielectric coefficient)A

% Considering a center frequency of 150 MHz, the lengths of the copper tubes of the balun corresponds to ~ lambda/4.

\subsubsection{Receiver Architecture and SDR Integration}

The receiver chain is built around commercially available SDR hardware. The primary units evaluated in this study were the LimeNet Mini, Ettus E310, and USRP 2920. These were selected for their support of GNU Radio integration, their capacity to digitize signals within the 100–200 MHz band, and their configurability in terms of gain, bandwidth, and sampling parameters. Each SDR was tested under controlled signal injection to characterize its linearity, sensitivity, and spectral response.\\

The analog front end includes one or two stages of low-noise amplification, depending on configuration, with optional filtering. Tests were performed using commercially available Mini-Circuits LNAs (ZX60-P103LN+) and bias tees. Additional filtering stages are being developed to reduce out-of-band RFI and suppress system gain variation near the band edges. Different gain settings were explored to assess the interplay between amplifier noise temperature, SDR quantization effects, and system linearity.

\subsection{Signal Processing and Software Architecture}

The digital processing pipeline was implemented using GNU Radio. A base flowgraph was constructed with SDR source blocks, configurable gain and bandwidth parameters, and a file sink for raw data output. Each acquisition consists of a set of overlapping frequency windows, each spanning 16 MHz, stitched together to cover the full 100–212 MHz band. The overlap between adjacent windows allows for the detection and removal of edge artifacts due to digital filtering.\\

Python scripts were written to parse the raw binary output, assign frequencies, and construct power spectra for each band segment. The software supports flexible gain averaging and FFT configurations. Optional real-time smoothing using Savitzky–Golay filters was implemented as a testbed for future RFI rejection and dynamic compression strategies. Although the data processing was performed offline for the results presented here, the pipeline supports near-real-time visualization, useful for system monitoring and on-site deployment.\\

Preliminary data are stored in structured directories organized by acquisition date and SDR type. The system is compatible with both Linux-based embedded platforms (as in the case of the Ettus E310 and LimeNet Mini) and external laptops interfaced via USB or Ethernet. All acquisition routines were designed to be portable and easily adapted to other SDR models or front-end configurations\footnote {The algorithms were developed by Ronal Sebastian Buitrago Parra, Yeison Andrés Quiroga Malambo, Brayan Nicolas Suarez Mongui and Felipe P. Mosquera.}.  \\

Two software packages were built, one for use with the Limenet mini (\texttt{lime\_eor}) and the other for the Ettus E310 (\texttt{etus\_eor}). Each main folder contains the same file structure but is configured for each SDR. Taking the Ettus folder as an example, it contains three files with .py extension and two folders, "data" and "graphics". 

\begin{figure}[h]
    \centering

    \includegraphics[scale=0.65]{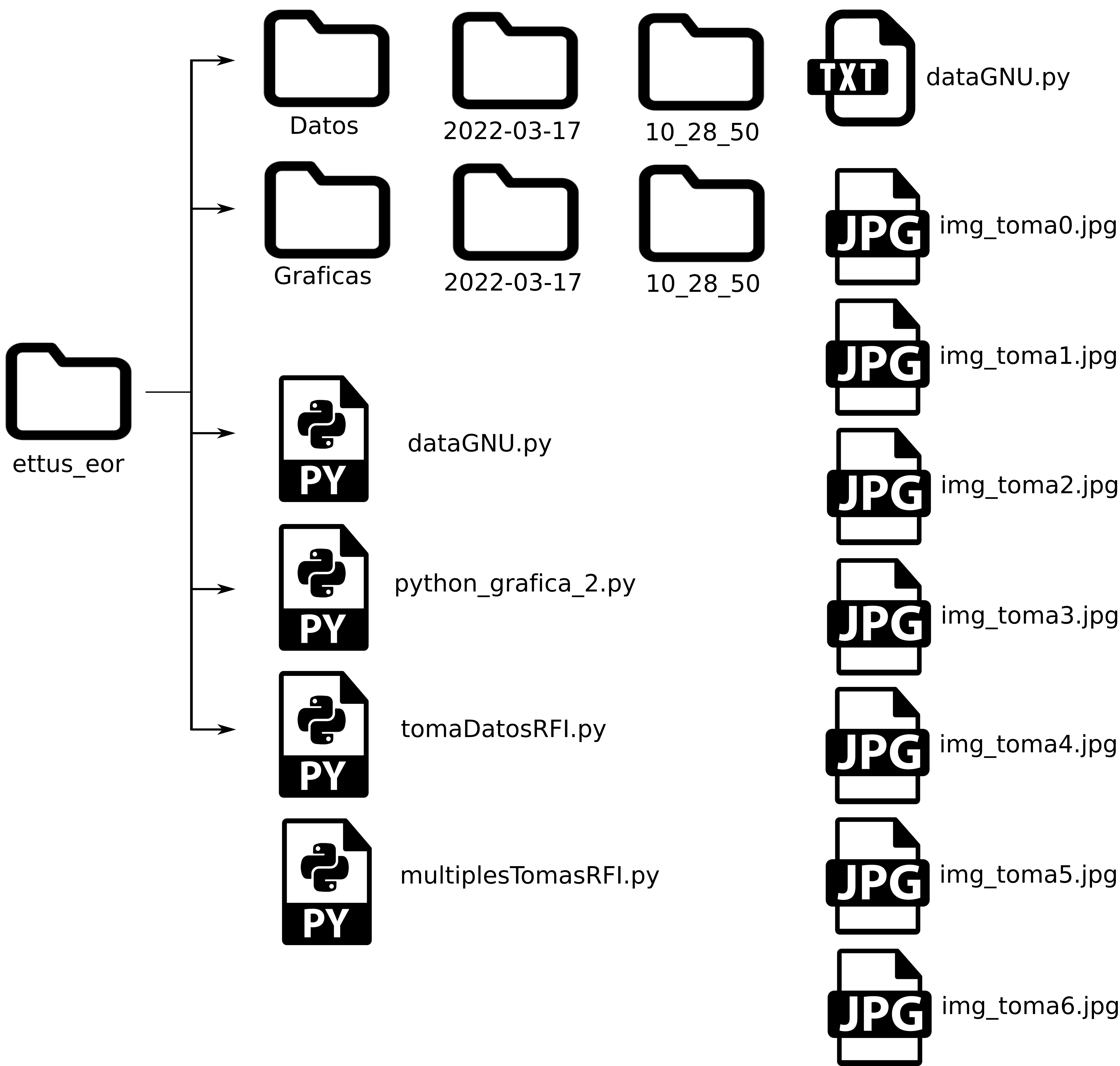}
    \caption{Software architecture using GNU Radio.} 
    \label{fig:med_1210}
\end{figure}

The data acquisition and processing workflow relies on a set of Python scripts developed around GNU Radio's flowgraph outputs. Figure \ref{fig:med_1210} shows a diagram of the implemented software architecture. Three main scripts structure the pipeline:

\begin{itemize}
    \item \texttt{python\_graficas\_2.py} is the auto-generated script produced by GNU Radio based on the implemented flowgraph. It defines the SDR configuration, sampling parameters, and data sinks for binary output.

    \item \texttt{dataGNU.py} processes the binary output produced by the SDR. It extracts the center frequency (typically 108 MHz) and defines the sampling rate (16~MS/s), resulting in a 16~MHz display bandwidth with an FFT size of 6144. The script parses the raw \texttt{.bin} file using \texttt{numpy}, reshapes the data into power spectra, and generates frequency-calibrated plots. Frequency axes, power scaling, and display parameters are set programmatically, enabling batch visualization of spectral segments.

    \item \texttt{tomaDatosRFI.py} automates the full-band acquisition by iterating over six center frequencies to cover the 100--212~MHz range in overlapping 16~MHz windows. For each window, the script generates spectrum plots and stores them in timestamped subdirectories. It also creates a \texttt{.txt} file with the raw numerical data for post-processing. At the end of each run, the center frequency is reset to 108~MHz for consistency. A companion script, \texttt{multiplesTomasRFI.py}, allows for repeated acquisitions over configurable time intervals for monitoring RFI variability.
\end{itemize}

\subsection{Antenna simulations}

We performed electromagnetic simulations of the antenna using Ansys HFSS to characterize its reflection coefficient ($S_{11}$) and beam chromaticity across the 100–200~MHz band. The model geometry matched the constructed blade dipole described in the previous section, including the feed gap, plate dimensions, and elevation above the ground plane. Chromaticity was evaluated in the principal planes $\phi=0^\circ$ and $\phi=90^\circ$, consistent with the expected sky coverage.\\

To reduce computational complexity, the simulated ground plane was limited to the central 1.3~m~$\times$~1.0~m aluminum sheet, excluding the full 5~m~$\times$~5~m wire mesh used in the physical system. This simplification may introduce minor discrepancies in the low-frequency edge of the band, where the extent of the ground plane has greater impact on impedance and beam structure.\\

The simulations were used to guide balun tuner optimization and to establish baseline expectations for matching and spectral smoothness, informing both mechanical construction and post-fabrication calibration strategies.\\

%\noindent Simulations were initially performed for the antennas without balun both 1:1 and 1:10 and later for the antennas with balun 1:1 and 1:10.

\section{Results}

This section presents the system-level performance metrics derived from the simulated and measured components, including antenna $S$-parameters, chromaticity, effective antenna area, source brightness temperature, and total system noise temperature under various preamplifier configurations. We also present the experimental validation of the antenna and receiver subsystems, including in-situ measurements of the antenna impedance response, laboratory tests of SDR performance, and gain characterization of the low-noise amplifiers (LNAs). Finally we evaluate the response of the SDR receivers in terms of linearity and noise levels.

\subsection{$S_{11}$ Response}

\begin{figure}[h]
    \centering
    %\caption{$S_1_1$ simulation (antenna 1:1) }
%    \includegraphics[scale=0.7]{Images/sim2/comp_sim_1_10balun_$S_1_1$_22_5cm.png}
    \includegraphics[scale=0.55]{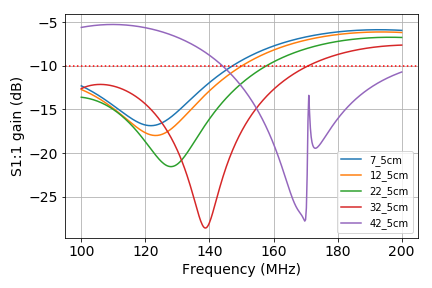} 
    \caption{$S_{11}$ simulations for the blade antenna considering 5 balun heights. }
    \label{fig:sim_100}
    \raggedright
    %\figurenote{\textit{Note.} Comparative of $S_1_1$ simulations for the 1:1 antenna considering 5 balun heights. Ansys HFSS license from Universidad de Chile.}
\end{figure}

The reflection coefficient $S_{11}$ was simulated for five different balun tuner heights, varying the effective electrical length of the antenna feed. These heights, ranging from 7.5 to 42.5~cm above the balun base, were selected to explore tuning effects across the 100--200~MHz band (see Table \ref{table:simbalun1}). The frequency range where $S_{11} < -10$~dB was taken as the operational bandwidth in each case. Figure \ref{fig:sim_100} illustrates the behavior of the blade antenna iterating between different heights of the balun tuner. \\

The results show that lower balun heights favor a flatter response near 100~MHz, while higher heights shift the resonance toward the upper band edge. The widest matched bandwidth was obtained at 32.5~cm, with $S_{11} < -10$~dB across 100--170~MHz. Peak reflection minima reached values below $-25$~dB in several configurations, indicating good impedance match in those regions. These trends informed the selection of tuner heights during measurements (see Section~5).

\begin{table}[h]
\centering
\caption{Blade antenna $S_{11}$ parameter simulation results.}
\label{table:simbalun1}
\begin{tabular}{lccccc}
\hline
\multicolumn{1}{c}{\textbf{Quantity}}    & \textbf{7.5cm} & \textbf{12.5cm} & \textbf{22.5cm} & \textbf{32.5cm} & \textbf{45.5cm} \\ \hline
Minimum frequency @-10dB (MHz)      & 100            & 100             & 100             & 100             & 145             \\
 
% Measurement min. freq. @-10dB (MHz)     & 112            & 112             & 111             & 111             & 117             \\
% 
Maximum frequency @-10dB (MHz)      & 147            & 150             & 157             & 170             & 200             \\
 
% Measurement max. freq. @-10dB (MHz)     & 145            & 149             & 158             & 166             & 179             \\
% 
Bandwidth (BW) (MHz)         & 447             & 50              & 57              & 70              & 55              \\
 
% Measurement BW (bandwidth) (MHz)        & 33             & 37              & 47              & 55              & 62              \\
% 
Minimum gain (dB)               & -16.85         & -17.98          & -21.25          & -28.54          & -27.24          \\
 
% Measurement min. gain (dB)              & -15.81         & -16.50          & -19.72          & -41.91          & -28.45          \\
% 
Frequency @ min. gain (MHz)  & 122            & 123             & 130             & 139             & 169             \\
 
% Measurement frequency @ min. gain (MHz) & 126            & 124             & 124             & 126             & 161            
\end{tabular} 

% \raggedright
% \figurenote{\textit{Note.} Blade antenna $S_1_1$ simulations for each balun tuner position.}
\end{table}

\subsection{Beam Chromaticity}
 
\begin{figure}[h]
    \centering
    \begin{minipage}{0.48\textwidth}
        \centering
        \includegraphics[width=\linewidth]{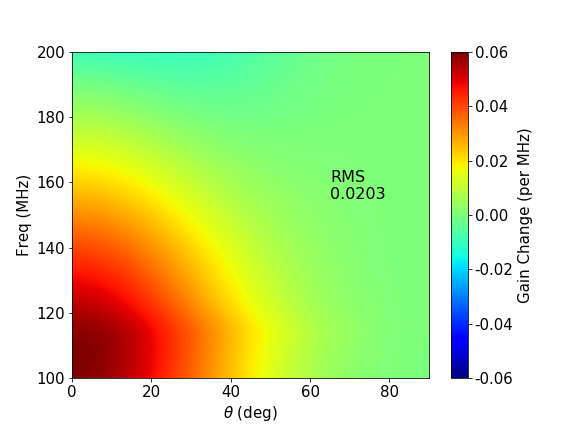}
    \end{minipage}
    \hfill
    \begin{minipage}{0.48\textwidth}
        \centering
        \includegraphics[width=\linewidth]{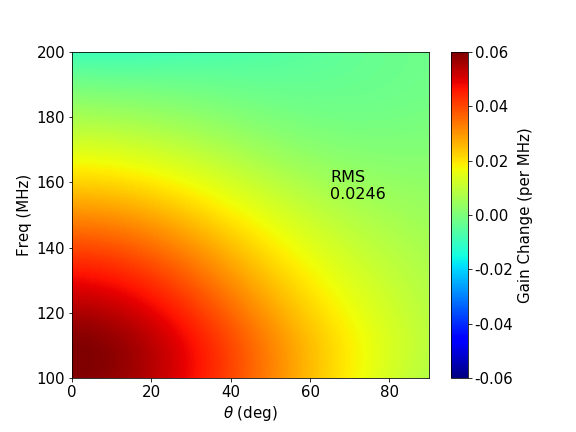}
    \end{minipage}
    \caption{Blade antenna chromaticity simulation for a balun height @ 22.5 cm. Gain change for the $\phi = 0^\circ$ (left) and $\phi = 90^\circ$ (right) planes.}
    \label{fig:sim_400}
    %\figurenote{\textit{Note.} Blade antenna chromaticity @ 22.5 cm. Gain change for planes $\phi = 0^\circ$ (left) and $\phi = 90^\circ$ (right). Ansys HFSS license from Universidad de Chile.}
\end{figure}

To assess beam chromaticity, far-field gain patterns were simulated as a function of frequency at a fixed balun height of 22.5~cm. Cuts at $\phi=0^\circ$ and $\phi=90^\circ$ were extracted over the 100--200~MHz range. Results of the chromaticity simulations for the blade antenna (with balun) are presented in Figure \ref{fig:sim_400}.  The chromatic response was characterized by tracking the gain variation at each elevation angle as a function of frequency and calculating the differential gain relative to a fixed reference frequency.\\

In both planes, the antenna exhibits reduced chromatic variation at zenith and increased variation near the horizon, as expected. For $\phi=0^\circ$, the gain variation below 20$^\circ$ elevation exceeds 3~dB over the band, while at higher elevations it remains relatively flat. For $\phi=90^\circ$, the behavior is broadly similar, with the smoothest region occurring between 60$^\circ$ and zenith. These results are consistent with expectations for a blade dipole and suggest that the antenna is sufficiently stable for beam modeling and calibration, especially if zenith-pointing observations are prioritized \citep{restrepo}.\\

Table~\ref{tab:sim_summary} summarizes the simulated bandwidth and reflection minima for each balun tuner height. While the simulations neglect ground screen extensions and lossy soil effects, they provide a useful reference for interpreting the measured response and guiding balun adjustments during field deployment.

\begin{table}[h]
\centering
\caption{Simulated $S_{11}$ parameters for different balun heights.}
\label{tab:sim_summary}
\begin{tabular}{lccccc}
\hline

\textbf{Parameter} & \textbf{7.5 cm} & \textbf{12.5 cm} & \textbf{22.5 cm} & \textbf{32.5 cm} & \textbf{42.5 cm} \\
\hline

Min freq. @ $S_{11} < -10$ dB (MHz) & 100 & 100 & 100 & 100 & 145 \\
 
Max freq. @ $S_{11} < -10$ dB (MHz) & 147 & 150 & 157 & 170 & 200 \\

Bandwidth (MHz)                    & 47  & 50  & 57  & 70  & 55  \\
 
Min $S_{11}$ (dB)                  & $-16.85$ & $-17.98$ & $-21.25$ & $-28.54$ & $-27.24$ \\

Freq. @ Min $S_{11}$ (MHz)         & 122 & 123 & 130 & 139 & 169 \\
\hline
\end{tabular}

\end{table}

\subsection{System Characterization Measurements}

We conducted system characterization measurements with the goal of quantifying system behavior across the 100--200~MHz band and identifying configurations that minimize instrumental contributions to the overall system temperature.

\subsubsection{Antenna Reflection Measurements}

The antenna reflection coefficient $S_{11}$ was measured on-site at the Páramo de Berlín using a Keysight FieldFox N9914A vector network analyzer (VNA) operating from 30~kHz to 6.5~GHz. As in the simulations, five balun tuner heights were tested to evaluate the impact of electrical length on the impedance match. Figure~\ref{fig:ant_1} shows the antenna design.\\

\begin{figure}[h]
    \centering
    %\caption{S11 measurement (antenna 1:1)} 
    \includegraphics[scale=0.58]{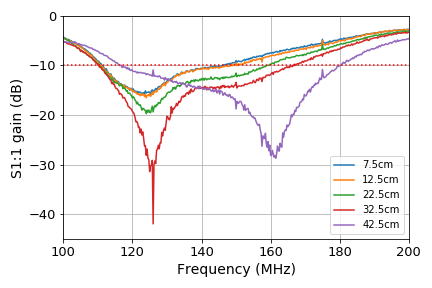} 
    \caption{$S_{11}$ VNA measurements considering different balun tuner heights.} 
    \raggedright
    %\figurenote{\textit{Note.} 1:1 antenna S11 parameter VNA measurements considering different balun tuner heights.}
    \label{fig:med_10}
\end{figure}

Figure~\ref{fig:med_10} shows the measured $S_{11}$ curves for the different balun heights. The results are summarized in Table~\ref{tab:s11_meas}, which includes the matched frequency range ($S_{11} < -10$~dB), total bandwidth, minimum reflection, and the frequency of minimum return loss.

\begin{table}[h]
\centering
\caption{Measured $S_{11}$ parameters for different balun tuner heights.}
\label{tab:s11_meas}
\begin{tabular}{lccccc}
\hline

\textbf{Parameter} & \textbf{7.5 cm} & \textbf{12.5 cm} & \textbf{22.5 cm} & \textbf{32.5 cm} & \textbf{45.5 cm} \\
\hline

Min freq. @ $S_{11} < -10$ dB (MHz) & 112 & 112 & 111 & 111 & 117 \\

Max freq. @ $S_{11} < -10$ dB (MHz) & 145 & 149 & 158 & 166 & 179 \\

Bandwidth (MHz)                    & 33  & 37  & 47  & 55  & 62  \\

Min $S_{11}$ (dB)                  & $-15.81$ & $-16.50$ & $-19.72$ & $-41.91$ & $-28.45$ \\

Freq. @ Min $S_{11}$ (MHz)         & 126 & 124 & 124 & 126 & 161 \\
\hline
\end{tabular}

\end{table}

The measured bandwidths and reflection minima show good agreement with the HFSS simulations, with minor shifts attributable to real ground conditions and the extended mesh ground plane not included in the simulation model. The deepest match was achieved at 32.5~cm, consistent with simulation predictions.

\subsubsection{SDR Linearity and Sensitivity}

\begin{figure}[h]
    \centering
    %\caption{Setups for SDR measurements.} 
    \includegraphics[scale=0.7]{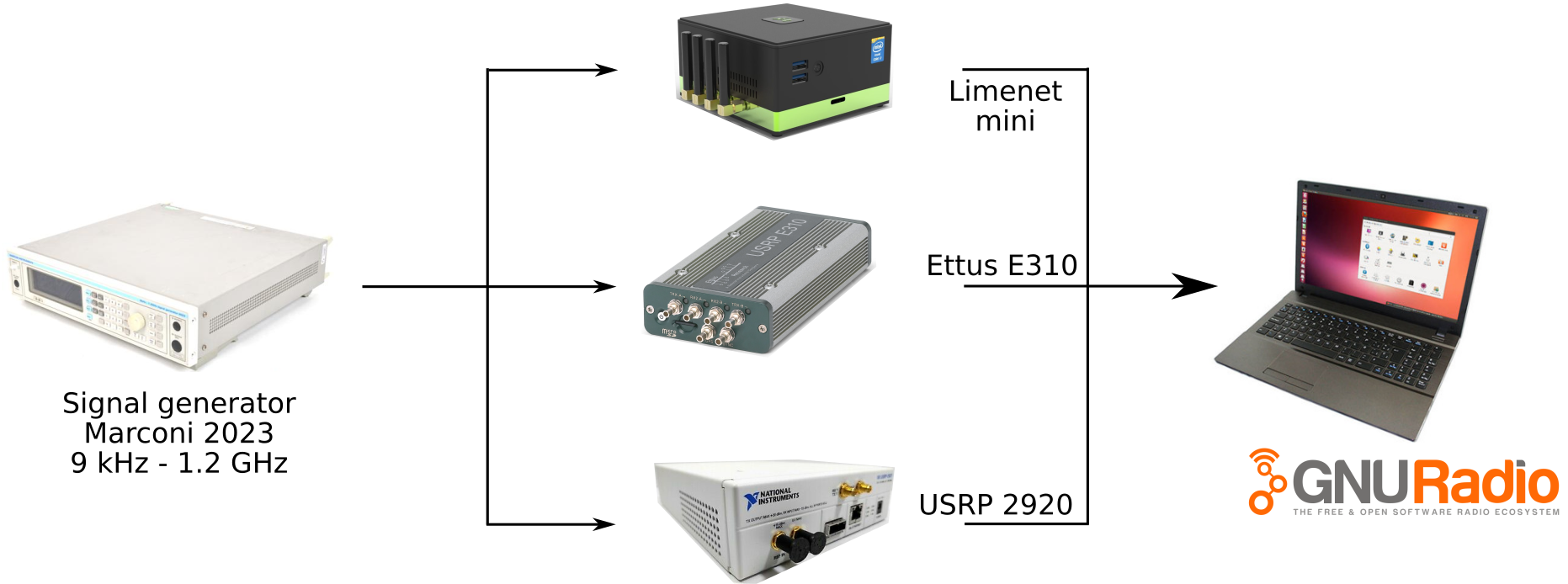} 
    \caption{Setup for SDR measurements.} 
    \label{fig:med_120}
    \raggedright
    %\figurenote{\textit{Note.} Images: www.ettus.com/all-products/e310/, www.ebay.com/itm/282506896648, www.artisantg.com/TestMeasurement/69602-1/Cobham-Aeroflex-Marconi-2023A-9-kHz-to-1-2-GHz-Signal-Generator, www.limemicro.com and www.gnuradio.org}
\end{figure}

To evaluate the dynamic range and sensitivity of the SDR receivers, we performed controlled signal injection tests using a calibrated RF signal generator (Marconi 2023A, maximum additional uncertainty in standard configuration: 0.5 dB) and a spectrum analyzer (R\&S ZVL6) (level measurement uncertainty: $<$0.5 dB). Figure~\ref{fig:med_120} shows the SDR measurement setup. Three SDRs were tested: LimeNet Mini, Ettus E310, and USRP 2920. Two measurement sets were conducted: one probing linearity across frequency at fixed input power, and another probing sensitivity at fixed frequency with varying input level.\\

For the linearity test, a $-30$~dBm tone was swept from 100 to 200~MHz in 1~MHz steps. The SDRs were configured with zero gain to reflect default field operation. All devices exhibited monotonic response across the band, though with different overall gain scaling. The Ettus E310 showed the most stable and consistent output, while the LimeNet Mini exhibited higher variability and a significantly elevated noise floor.\\

\begin{figure}[h]
    \centering
    %\caption{SDRs linearity response} 
    \includegraphics[scale=0.58]{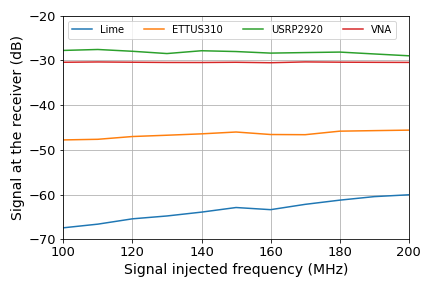}
    \caption{SDRs linearity response. Limenet mini, Ettus E310 and USRP2920 linearity response @ gain=0, injecting -30 dBm and varying the frequency between 100 and 200 MHz.}
    \label{fig:med_12}
    \raggedright
    %\figurenote{\textit{Note.} Limenet mini, Ettus E310 and USRP2920 linearity response @ gain=0, injecting -30dBm and varying the frequency between 100 and 200 MHz.}
\end{figure}

\begin{figure}[h]
    \centering

    % Fila superior: 2 imágenes lado a lado
    \begin{minipage}{0.48\textwidth}
        \centering
        \includegraphics[width=\linewidth]{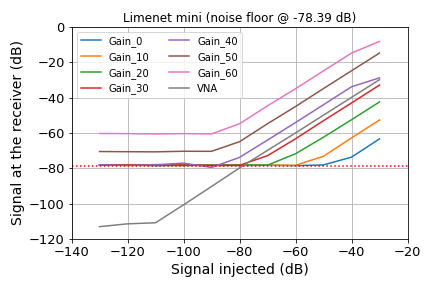}
    \end{minipage}
    \hfill
    \begin{minipage}{0.48\textwidth}
        \centering
        \includegraphics[width=\linewidth]{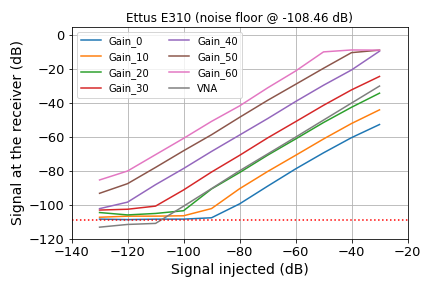}
    \end{minipage}

    % Espacio entre filas
    \vspace{0.5em}

    % Fila inferior: imagen centrada
    \begin{minipage}{0.48\textwidth}
        \centering
        \includegraphics[width=\linewidth]{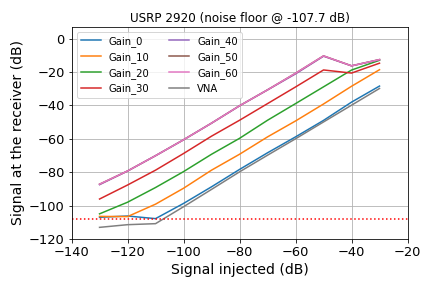}
    \end{minipage}

    \caption{SDRs sensitivity response. Limenet mini (top-left), Ettus E310 (top-right) and USRP2920 (bottom) sensitivity response @ SDR gain = 0, injecting 150 MHz and varying the RF level between $-$30 and $-$120 dBm.}
    \label{fig:med_14}
    %\figurenote{\textit{Note.} Limenet mini (top-left), Ettus E310 (top-right) and USRP2920 (bottom) sensitivity response @ SDR gain = 0, injecting 150 MHz and varying the RF level between $-$30 and $-$120 dBm.}
\end{figure}

For sensitivity, a 150~MHz tone was injected at power levels from $-30$ to $-130$~dBm. Each SDR was tested across gain settings from 0 to 60~dB. The LimeNet Mini showed detection thresholds around $-60$~dBm at 0~dB gain, improving to about $-90$~dBm at 30~dB. The E310 demonstrated lower detection thresholds across the board, reliably detecting signals as low as $-110$~dBm at moderate gain settings. The USRP 2920 offered the best raw sensitivity but showed compression artifacts at high gain. Figures \ref{fig:med_12}, \ref{fig:med_14} show the results of these measurements. \\

These results indicate that while all three SDRs are suitable for exploratory measurements, careful gain calibration and noise floor characterization are essential, especially when seeking stability at millikelvin scales.

\subsubsection{LNA Gain Characterization}

\begin{figure}[h]
    \centering
    %\caption{Pre-amplifiers LNA $S_2_1$ response}
    \includegraphics[scale=0.52]{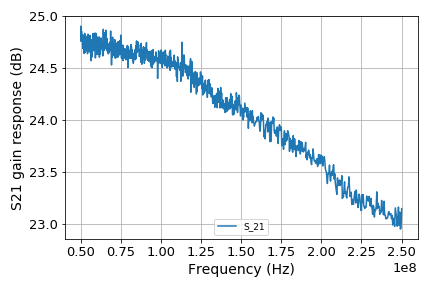}
    \includegraphics[scale=0.52]{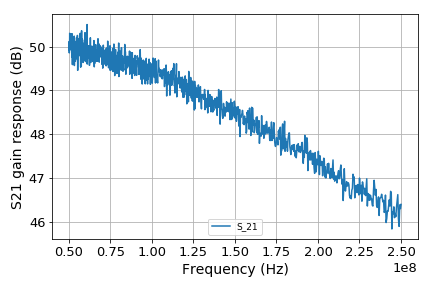} 
    \caption{Pre-amplifiers LNA $S_{21}$ response. $S_{21}$ response of the current pre-amplifier LNA ZX60-P103LN+, testing one LNA (top) and testing two-serial LNAs (bottom).}
    \raggedright
    %\figurenote{\textit{Note.} $S_2_1$ response of the current pre-amplifier LNA ZX60-P103LN+, testing one LNA (top) and testing two-serial LNAs (bottom).}
    \label{fig:med_1200}
\end{figure}

The analog front-end amplification chain employs Mini-Circuits ZX60-P103LN+ LNAs. To quantify gain and spectral flatness, we measured the forward transmission parameter ($S_{21}$) of both single and cascaded LNA configurations using the FieldFox VNA. Figure \ref{fig:med_1200} shows the LNA response (level measurement uncertainty: $<$0.5 dB).\\

A single amplifier exhibited flat gain of approximately 24.5~dB from 50 to 100~MHz, decreasing gradually to 23.0~dB at 200~MHz. The cascaded pair delivered a combined gain of 49.5~dB at 100~MHz, tapering to about 47.6~dB at 200~MHz. This decline reflects the specified performance and is consistent with the manufacturer's datasheet.\\

The measurements confirm that the amplifier stages introduce minimal spectral structure over the band of interest and are suitable for use with moderate gain SDRs. Nonetheless, future configurations will incorporate bandpass filters to suppress out-of-band noise and mitigate gain compression near strong RFI sources.

\subsection{Antenna performance metrics}

\subsubsection{Effective Area and Source Temperature}

The antenna gain was estimated from HFSS simulations to be approximately 7~dBi at zenith near 150~MHz, corresponding to a linear gain of $G = 5.01$. The effective area $A_\mathrm{e}$ at this frequency is then given by

\begin{equation}
A_\mathrm{e} = \frac{\lambda^2 G}{4\pi} = \frac{(2\,\mathrm{m})^2 \cdot 5.01}{4\pi} \approx 1.59~\mathrm{m}^2,
\end{equation}

where $\lambda = c / \nu$ is the wavelength at 150~MHz.\\

To estimate the system response to a strong known radio source, we consider the quiet Sun, which has a flux density of $1.24 \times 10^6$~Jy at 150~MHz. The corresponding antenna temperature is

\begin{equation}
T_\mathrm{src} = \frac{A_\mathrm{e}}{2k} \cdot S = \frac{1.59}{2k} \cdot 1.24 \times 10^{-20}~\mathrm{W\,m^{-2}\,Hz^{-1}} \approx 716~\mathrm{K}.
\end{equation}

This value provides a consistency check for radiometric calibration and detection threshold estimates.

\subsubsection{System Temperature Estimates}

We estimate the total system temperature $T_\mathrm{sys}$ under three scenarios: (1) without a preamplifier, (2) using the current preamplifier (two cascaded ZX60-P103LN+ amplifiers), and (3) a proposed preamplifier chain including filtering stages. The Friis equation was used to compute the equivalent noise temperature of the system, incorporating cable losses, amplifier gains, and manufacturer-specified noise figures.\\

In all cases, the antenna temperature was assumed to be $T_\mathrm{ant} = 300$~K, based on ITU-R P.372-7 and accounting for both galactic noise and moderate terrestrial RFI contributions at 200~MHz.

\subsubsection{System Behavior without Preamplifier}

\begin{figure}[h]
    \centering
    %\caption{Front-end block diagram (without pre-amplifier)}
    \includegraphics[scale=0.8]{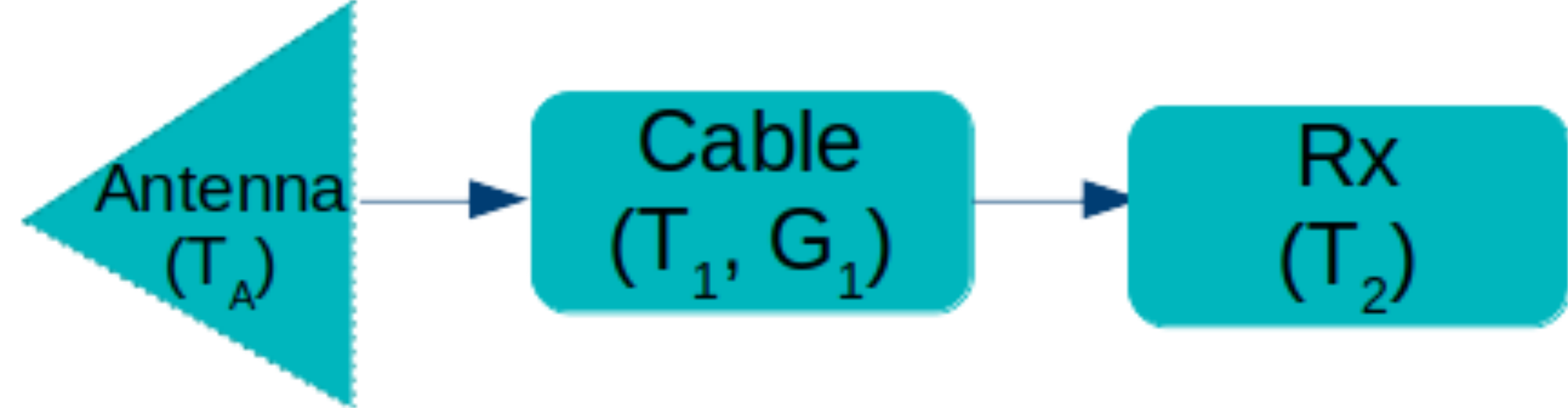} 
    \caption{Front-end block diagram (without pre-amplifier).}
    \raggedright
    %\figurenote{\textit{Note.} Diagram of the subsystems to calculate the overall temperature of the radio telescope without pre-amplifier.}
    \label{fig:temp_ambiente}
\end{figure}

Without any preamplification, the signal is passed through 30~m of LMR400 coaxial cable ($\sim$1.5~dB loss) to the SDR receiver. Figure \ref{fig:temp_ambiente} illustrates a conceptual diagram of the subsystems to calculate the overall temperature of the system. In this case, no pre-amplifier was considered. To determine the system temperature without a pre-amplifier, we considered the contributions of both the cable and the receiver. The 100-foot LMR400 cable exhibited a gain of -1.5 dB (linear gain of 0.993), a noise figure of 1.5 dB, and a temperature of 119.36 K. The Ettus E310 receiver had a noise figure of 8 dB and a temperature of 1539.776 K. For an Ettus E310 receiver with an 8~dB noise figure, the total system temperature is:

\begin{equation}
T_\mathrm{sys} = T_\mathrm{ant} + T_\mathrm{rx} \approx 300~\mathrm{K} + 1370~\mathrm{K} = 1670~\mathrm{K},
\end{equation}

indicating that cable attenuation and receiver noise dominate in this configuration.\\

%The Noise figure is obtained by using the Friss aproximation but considering the %noise figure instead of the temperature:
%$$
%NF_T_O_T_A_L=NF_1 + \frac{NF_2-1}{G_1 (linear)}
%$$
%$$
%NF_T_O_T_A_L= 3.5481 + \frac{1.2589-1}{0.2818} = 4.4668(noise factor) = 6.49633dB %(noise figure)
%$$
%where 4.4668 corresponds to the noise factor and 6.4999dB to the noise figure.
%In order to corroborate the equations using $T=t_0(NF-1)$ we obtain:
%$$
%T=t_0(NF-1) = 290(4.4668-1) = 1005,372 K \approx 1005,4155 K
%$$

Table \ref{table:result_calculos} presents the results of the calculations for the system, considering that $T_{REF}$ is the temperature of the receiver (all stages except the antenna), $T_{ANT}$ is the temperature of the antenna, $T_{SYS}$ is the temperature of the system, $NF$ the noise figure, and $F$ is the noise factor.

\begin{table}[h]
\centering
\caption{System temperature and noise figure results (without pre-amplifier)}
\label{table:result_calculos}
\begin{tabular}{cc}
\hline
\textbf{Parameter}       & \textbf{Value}             \\ \hline

Antenna Effective area   & 1.5947 m$^2$ \\
 
Source temperature (Sun) & 716.459 K                  \\

$T_\text{RX}$                 & 1669.99 K                \\
 
$T_\text{ANT} $        & 300 K                      \\

$T_\text{SYS}$           & 1969.99 K                \\
 
$NF$                       & 8.917 dB                   \\

$F$                        & 7.79                      
\end{tabular} 
 \\
\raggedright

\end{table}

\subsubsection{System Behavior With Current Preamplifier}

\begin{figure}[h]
    \centering
    %\caption{Front-end block diagram (current pre-amplifier)} 
    \includegraphics[scale=0.8]{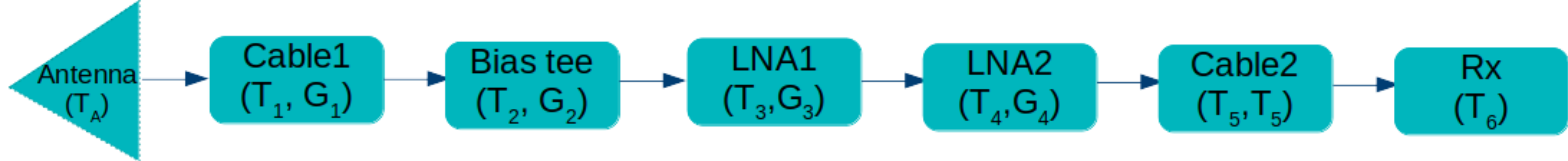} 
    \caption{Front-end block diagram (current pre-amplifier).} 
    \raggedright
    %\figurenote{\textit{Note.} Diagram of the subsystems to calculate the overall temperature of the radio telescope considering the current pre-amplifier.}
    \label{fig:preamp2}
\end{figure}

The existing analog front end includes two cascaded ZX60-P103LN+ amplifiers (each with 23~dB gain and 1.2~dB NF), along with bias tees and short interconnects. Figure \ref{fig:preamp2} presents a schematic of the current pre-amplifier stages and Table \ref{table:recom} illustrates the following parameters: component model, gain (linear and in dB), noise figure (dB), and temperature (K). Accounting for cable loss and gain distribution, the estimated system temperature reduces significantly:

\begin{equation}
T_\mathrm{sys} \approx 300~\mathrm{K} + 153~\mathrm{K} = 453~\mathrm{K},
\end{equation}

with an equivalent noise figure of approximately 4.1~dB.

\begin{table}[h]
\centering
\caption{System temperature estimation parameters (current pre-amplifier)}
\label{table:recom}
\begin{tabular}{cccccc}
\hline
\textbf{Component}    & \textbf{Reference}                    & \textbf{$G$(dB)} & \textbf{$G$(linear)} & \textbf{$NF$(dB)} & \textbf{$T$(K)} \\  \hline

Cable1 ($T_1$, $G_1$)   & 086-2SM+ $\times$ 2 in & -0.03          & 0.993              & 0.03            & 2.01          \\
 
Bias tee ($T_2$, $G_2$) & ZFBT-4R2GW-FT                         & -0.6           & 0.871              & 0.6             & 42.96         \\

LNA1 ($T_3$, $G_3$)     & ZX60-P103LN+                          & 23             & 199.526            & 1.2             & 92.92         \\
 
LNA2 ($T_4$, $G_4$)     & ZX60-P103LN+                          & 23             & 199.526            & 1.2             & 92.92         \\

Cable3 ($T_5$, $G_5$)   & LMR400 $\times$100 ft                  & -1.5           & 0.993              & 1.5             & 119.36        \\
 
Receiver ($T_6$)       & Ettus E310                                & --             & --                 & 8               & 1539.776     \\
\end{tabular} 
\\
\raggedright
\end{table}

Table \ref{table:result_calculos_amplificador} presents the results for the current pre-amplifier in terms of the $T_{RX}$, $T_{ANT}$, $T_{SYS}$, $NF$, and $F$.

\begin{table}[h]
\centering
\caption{System temperature and noise figure results (current pre-amplifier)}
\label{table:result_calculos_amplificador}
\begin{tabular}{cc}
\hline
\textbf{Parameter}       & \textbf{Value}             \\ \hline

Antenna Effective area   & 1.5947 m$^2$ \\
 
Source temperature (sun) & 716.459 K                  \\

$T_\text{RX}$                  & 153.27 K                \\
 
$T_\text{ANT}$               & 300 K                      \\

$T_\text{SYS}$               & 453.27 K                \\
 
$NF$                       & 4.09 dB                   \\

$F$                        & 2.56                      
\end{tabular}
\\
\raggedright

\end{table}

\subsubsection{System Behavior With Proposed Filtered Preamplifier}

\begin{figure}[h]
    \centering
    %\caption{Front-end block diagram (suggested pre-amplifier)}
    \includegraphics[scale=0.45]{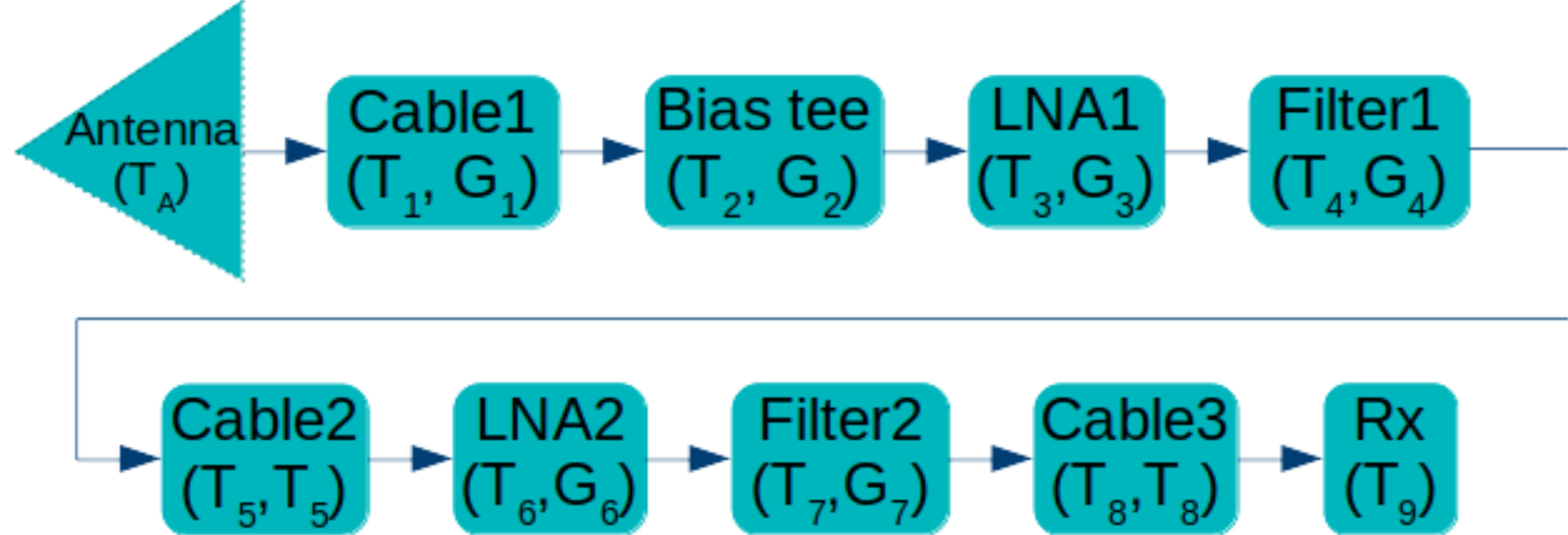} 
    \caption{Front-end block diagram (proposed pre-amplifier).}
    \label{fig:rec_diagram}
    \raggedright
    %\figurenote{\textit{Note.} Diagram of the subsystems to calculate the overall temperature of the radio telescope considering the suggested pre-amplifier.}
\end{figure}

The proposed configuration adds high- and low-pass filters to suppress out-of-band power and improve gain flatness. Two ZX60-33LNR-S+ amplifiers, each with 24.7~dB gain and 1.1~dB NF, are combined with Mini-Circuits SHP-20+ and SLP-250+ filters. These filters restrict the frequencies adjacent to those of interest (between 100 and 200 MHz) (see Figure \ref{fig:rec_diagram}). This architecture is based on the pre-amplification stage used in the MIST and MINI-MIST projects \citep{rest}. The resulting system temperature is further reduced:

\begin{equation}
T_\mathrm{sys} \approx 300~\mathrm{K} + 143~\mathrm{K} = 443~\mathrm{K},
\end{equation}

with an overall noise figure of 4.0~dB. This improvement is modest in absolute terms, but important for achieving higher dynamic range and minimizing foreground leakage in the final power spectrum. The values obtained for the system temperature estimation parameters are presented in Table \ref{table:recom2}. Table \ref{table:result_calculos_amplificador} shows the results of the calculations of the system temperature using a pre-amplifier.

\begin{table}[h]
\centering
\caption{System temperature estimation parameters (proposed pre-amplifier)}
\label{table:recom2}
\begin{tabular}{cccccc}
\hline
\textbf{Component}    & \textbf{Reference}     & \textbf{$G$(dB)} & \textbf{$G$(linear)} & \textbf{$NF$(dB)} & \textbf{$T$(K)} \\ \hline

Cable1 ($T_1, G_1$)   & 086-2SM+ $\times$ 2 in & -0.03          & 0.993              & 0.03            & 2.01          \\
 
Bias tee ($T_2, G_2$) & ZFBT-4R2GW-FT          & -0.6           & 0.871              & 0.6             & 42.96         \\

LNA1 ($T_3, G_3$)     & ZX60-33LNR-S+          & 24.7           & 295.121            & 1.1             & 83.592        \\
 
Filter1 ($T_4, G_4$)  & SHP-20+                & -0.13          & 0.971              & 0.13            & 8.812         \\

Cable2 ($T_5, G_5$)   & 086-2SM+ $\times$ 2 in & -0.03          & 0.993              & 0.03            & 2.01          \\
 
LNA2 ($T_6, G_6$)       & ZX60-33LNR-S+          & 24.7           & 295.121            & 1.1             & 83.592        \\

Filter2 ($T_7, G_7$)  & SLP-250+               & -0.27          & 0.94               & 0.27            & 18.601        \\
 
Cable3 ($T_8, G_8$)   & LMR400 $\times$100ft   & -1.5           & 0.993              & 1.5             & 119.36        \\

Receiver ($T_9$)       & Ettus E310             & --             & --                 & 8               & 1539.776     
\end{tabular} 
\\
    \raggedright
\end{table}

% Where $T_R_X$ is the temperature of the receiver (all stages except the antenna), $$T_\text{ANT}$$ is the temperature of the antenna, $$T_\text{SYS}$$ is the temperature of the system, $NF$ is the noise figure and $F$ is the noise factor. 

\begin{table}[h]
\centering
\caption{System temperature and noise figure results (proposed pre-amplifier)}
\label{table:result_calculos_amplificador2}
\begin{tabular}{cc}
\hline
\textbf{Parameter}       & \textbf{Value}             \\ \hline

Antenna Effective area   & 1.5947 m$^2$ \\
 
Source temperature (sun) & 716.459 K                  \\

$T_\text{RX}$                  & 143.32 K                \\
 
$T_\text{ANT}$               & 300 K                      \\

$T_\text{SYS}$               & 443.32 K                \\
 
$NF$                       & 4.02 dB                   \\

$F$                        & 2.53                      
\end{tabular} 
\\
\raggedright
\end{table}

\subsection{SDR Sensitivity and Linearity Comparison}

\begin{figure}[h]
    \centering
    %\caption{SDRs sensitivity comparisson}
    \includegraphics[scale=0.52]{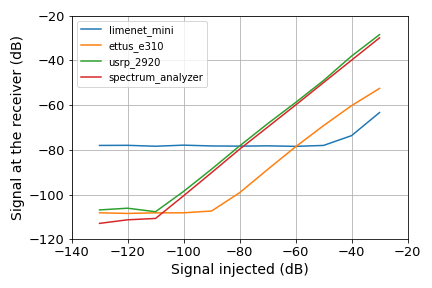}
    \includegraphics[scale=0.52]{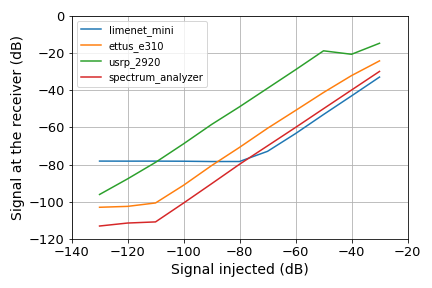} 
    \caption{SDRs sensitivity comparison: Limenet mini, Ettus E310 and USRP2920 @ gain = 0 dB (left) and @ gain = 30 dB (right), injecting 150 MHz and varying the RF level between $-$30 and $-$120 dBm.}
    \label{fig:med_304}
    \raggedright
    %\figurenote{\textit{Note.} Limenet mini, Ettus E310 and USRP2920 sensitivity comparison @ gain = 0 dB (left) and @ gain = 30 dB (right), injecting 150 MHz and varying the RF level between $-$30 and $-$120 dBm.}
\end{figure}

Three SDR receivers were evaluated using signal injection tests: the LimeNet Mini, Ettus E310, and USRP 2920. Each device showed distinct noise floors and gain characteristics, largely due to differences in ADC architecture and internal filtering. Figure~\ref{fig:med_304} shows the response of each device to a fixed 150MHz tone at varying RF input levels, measured at receiver gain settings of 0dB and 30~dB.

The Ettus E310 exhibited the lowest noise floor ($-108.5$~dB) and the most linear response, closely matching the reference spectrum analyzer. The USRP 2920 performed comparably in sensitivity but showed early onset of gain compression. The LimeNet Mini, although functional, had a significantly higher noise floor ($-78$~dB) and poorer spectral stability.

The optimal operating gain for each SDR was determined by sweeping gain settings from 0 to 60dB. For the LimeNet, the best tradeoff between distortion and sensitivity occurred at 30dB; for the E310, it was 20dB; and for the USRP, 0dB. These values provide guidance for system configuration during sky measurements and influence preamplifier design, particularly for minimizing total system temperature.

Tests involving fixed RF power and swept frequency confirmed that the USRP 2920 delivered the most spectrally flat and consistent output, especially at high input levels. The E310 showed slightly more variation but maintained acceptable linearity.

\subsection{System Integration and In-Situ Measurements}

The complete system (antenna, transmission line, preamplifier, and SDR) was deployed for initial sky testing at the Páramo de Berlín. The preamplifier followed the configuration discussed above, and the Ettus E310 was used as the receiver, configured with 0 dB gain. The software setup, including FFT and averaging parameters, followed the definitions in Section~3.3.\\

\begin{figure}[H]
    \centering
    %\caption{In-situ measurements} 
    \includegraphics[scale=0.24]{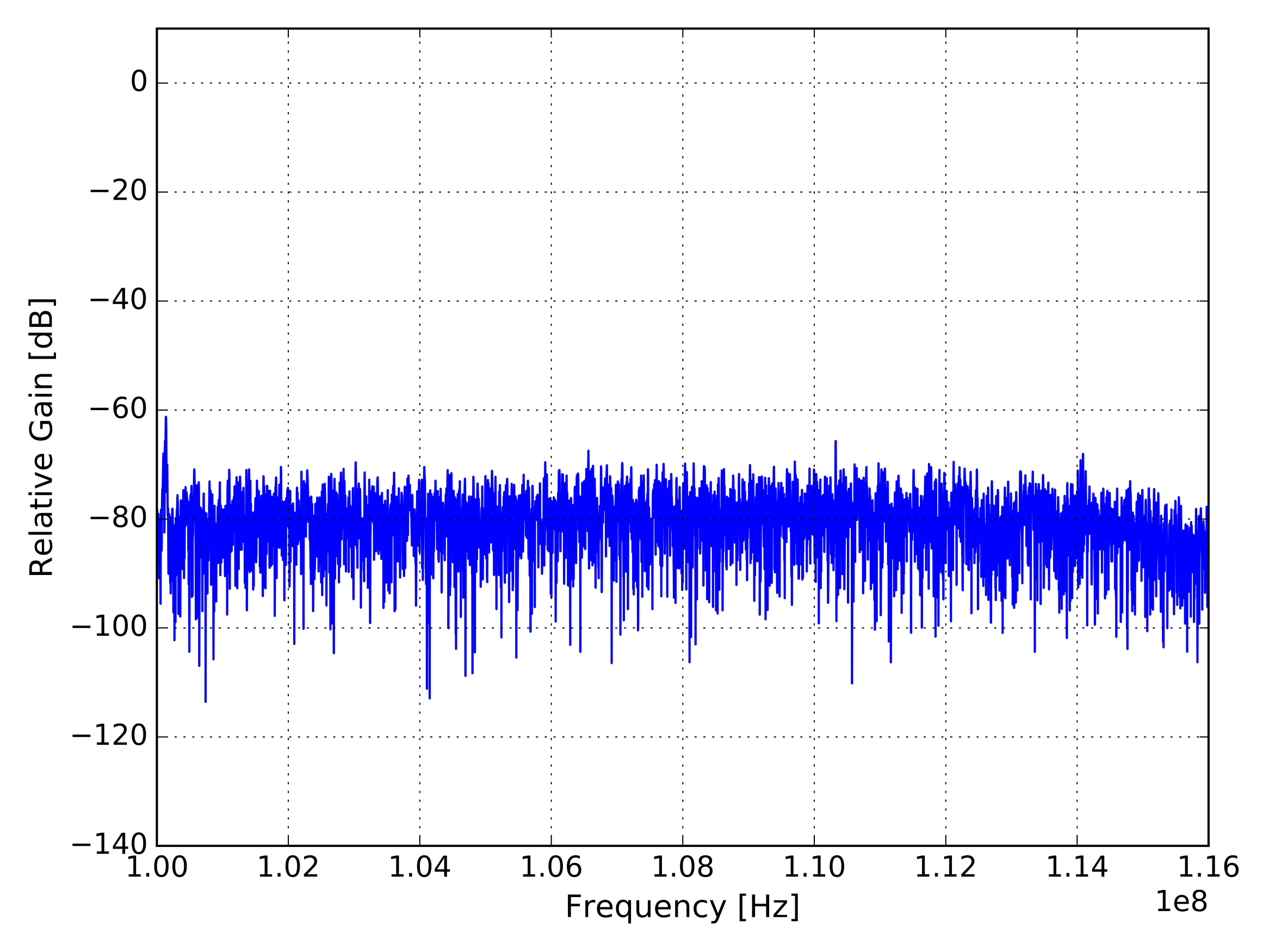}
    \includegraphics[scale=0.24]{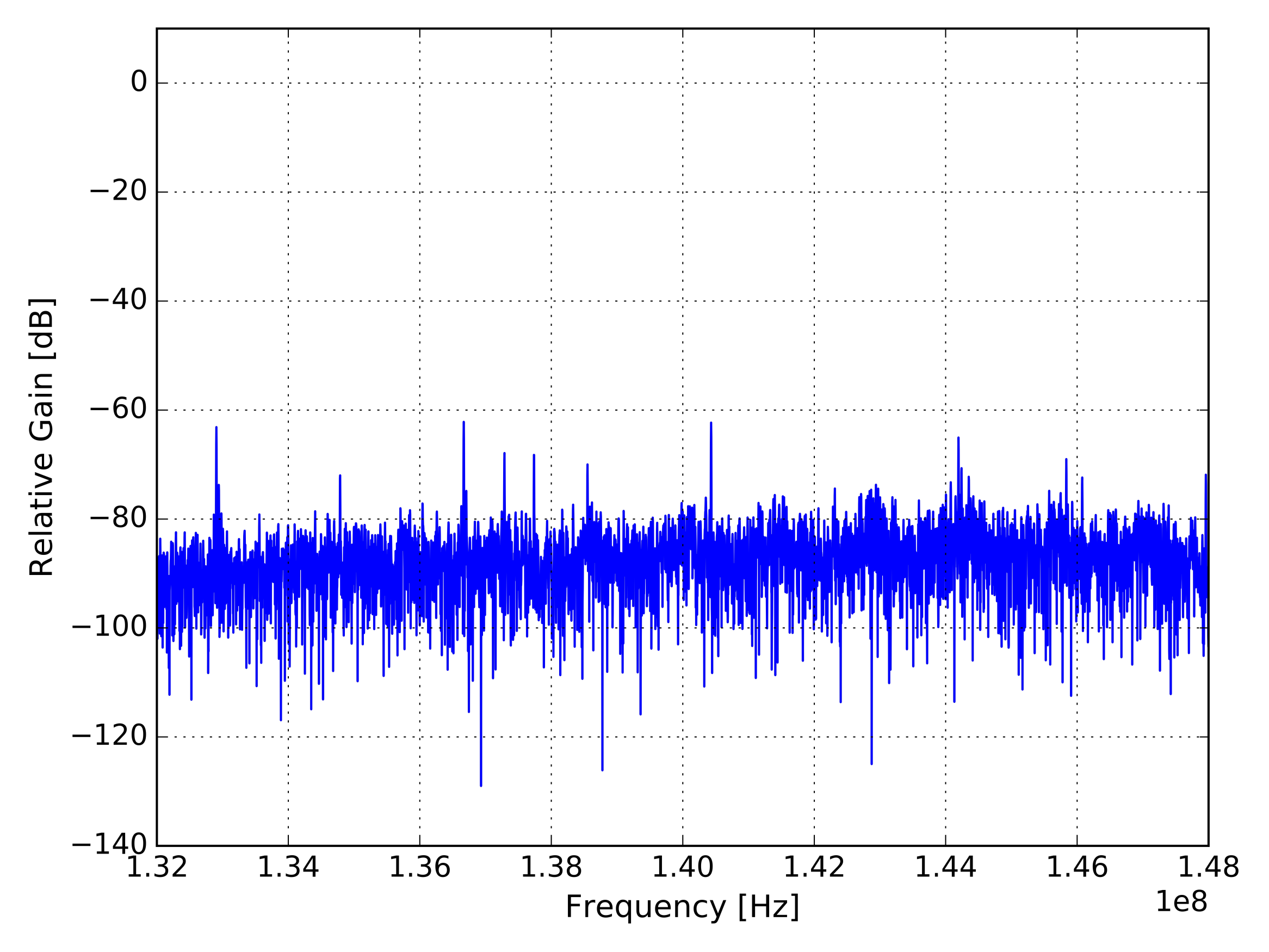}
    \includegraphics[scale=0.24]{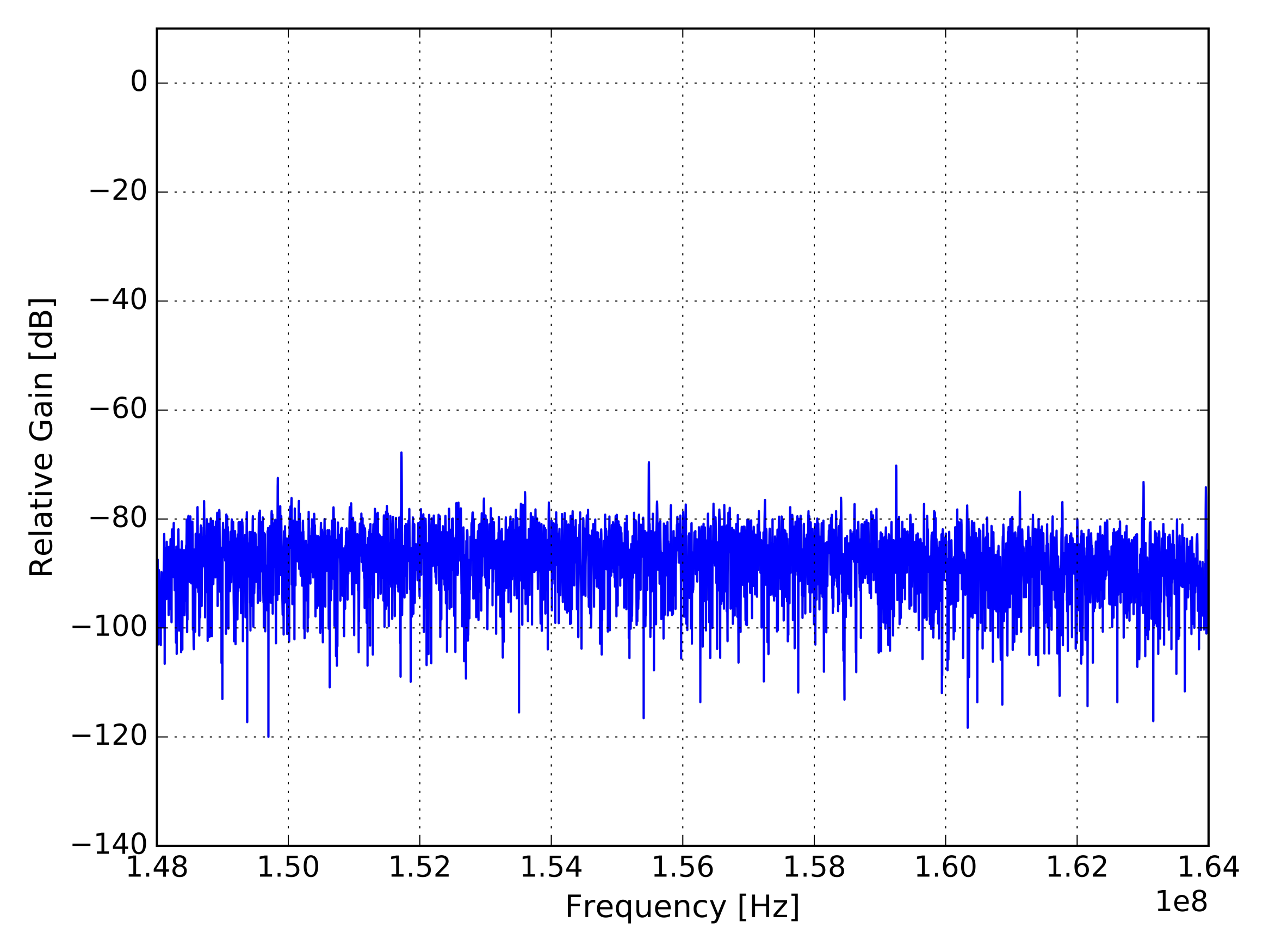}
    \includegraphics[scale=0.24]{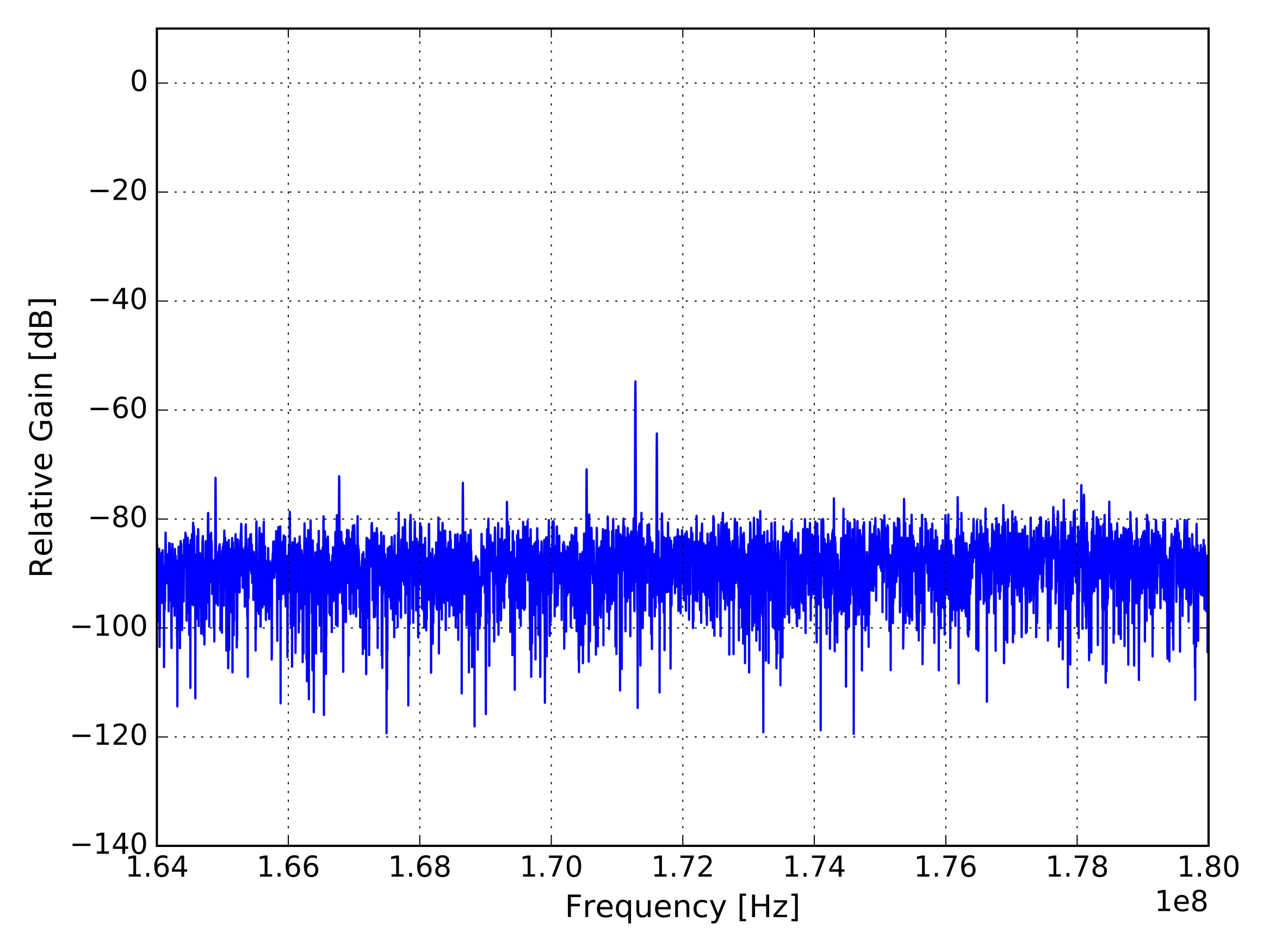}
    \includegraphics[scale=0.24]{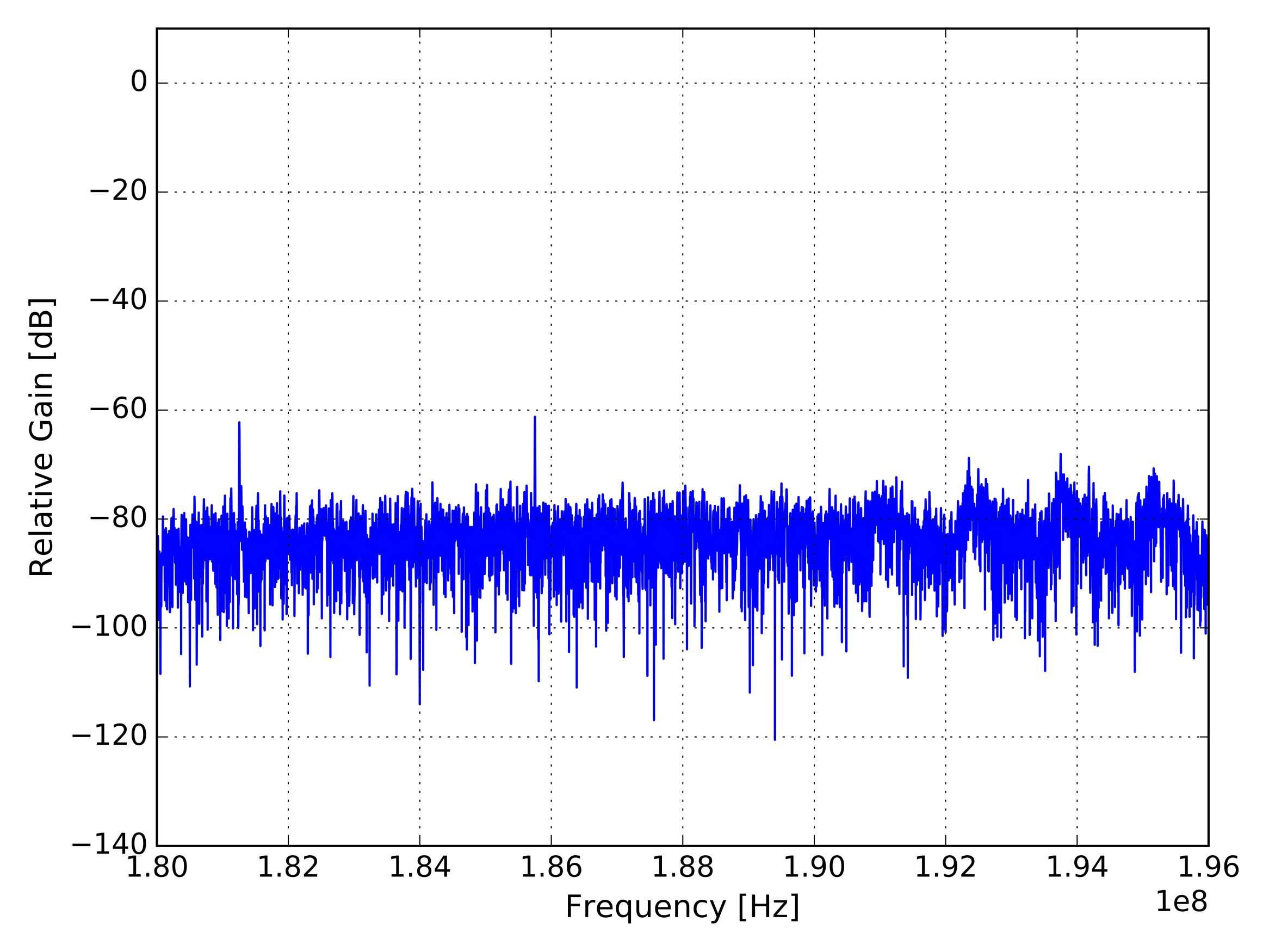}
    \includegraphics[scale=0.24]{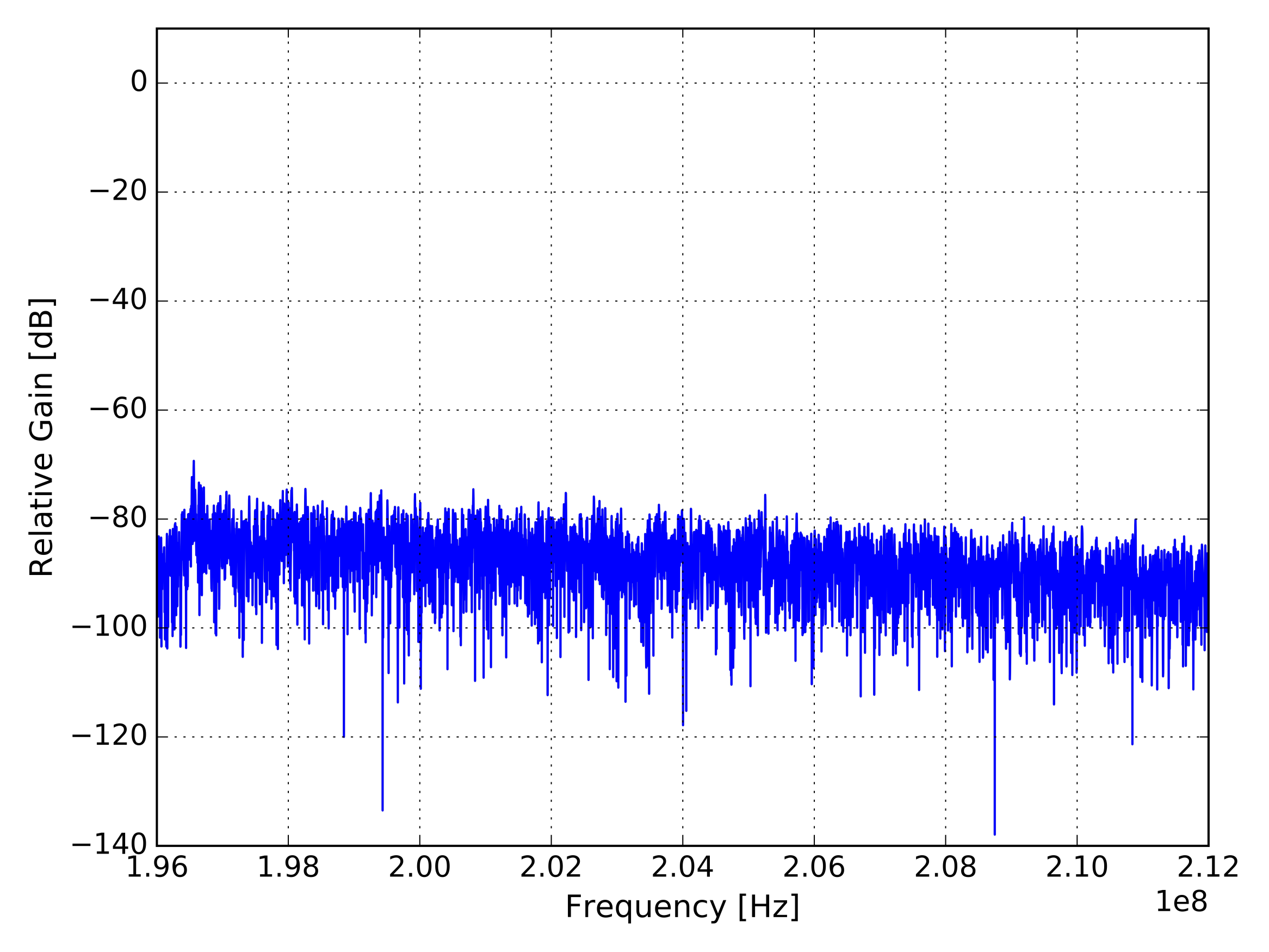} 
    \caption{In-situ RFI measurements considering the following frequency ranges: 100-116 MHz, 132-148 MHz, 148-164 MHz, 164-180 MHz, 180-196 MHz and 196-212 MHz. The 116-132 MHz measurements are presented in the following figure.} 
    \label{fig:med_501}
    \raggedright
    %\figurenote{\textit{Note.} In-situ measurements considering the following frequency ranges (top to bottom): 164-180 MHz, 180-196 MHz and 196-212 MHz.}
\end{figure}

Figure \ref{fig:med_501} show the measured spectra across the full 100--212MHz band, segmented into seven overlapping 16 MHz windows. The response follows expectations based on previous component-level tests. Signal and noise are amplified together, underscoring the need for filtering stages to suppress out-of-band contributions.\\

\begin{figure}[h]
    \centering
    %\caption{In-situ measurements (with averages)} 
    \includegraphics[scale=0.3]{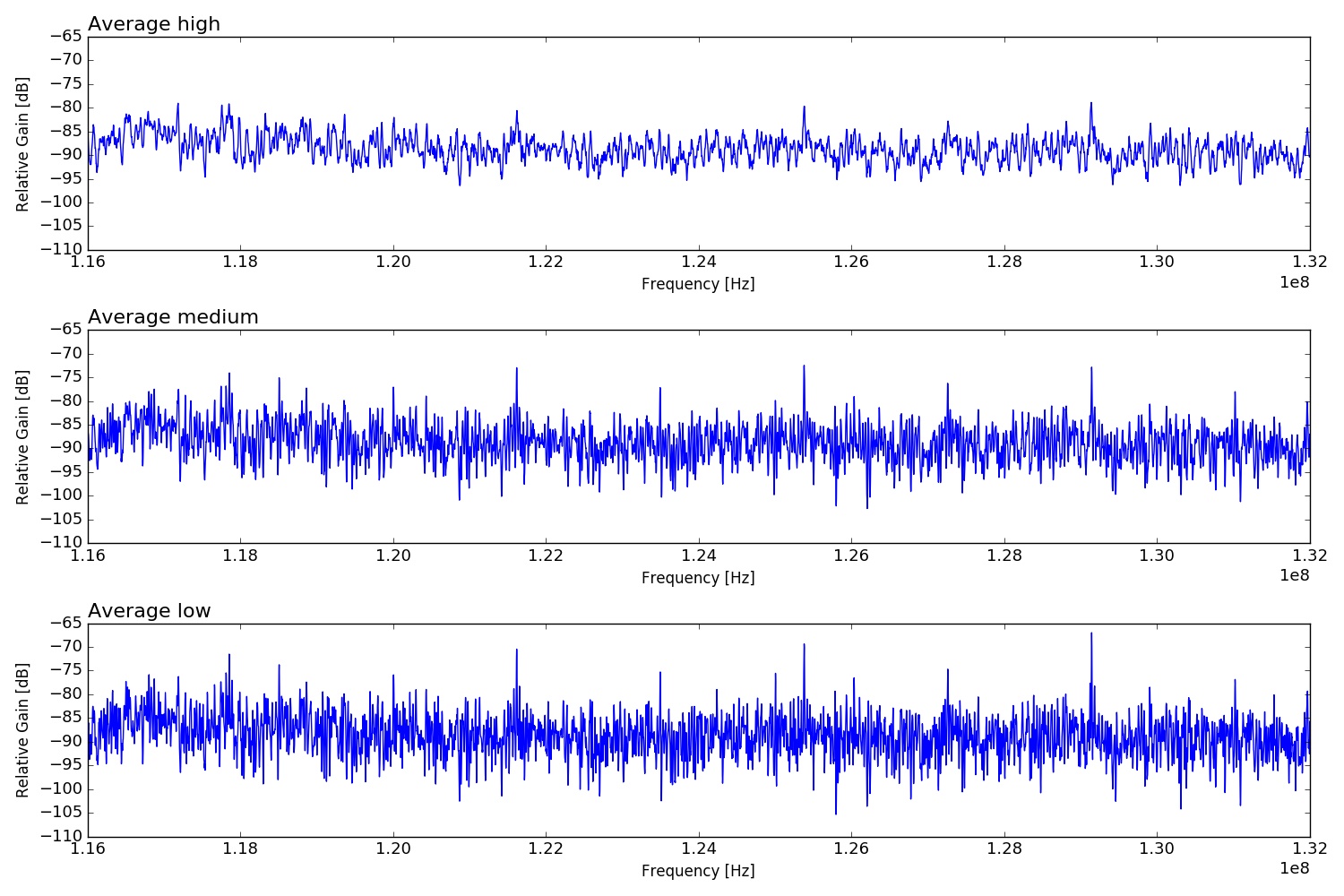} 
    \caption{In-situ RFI measurements with high averaging (top), medium averaging (middle) and low averaging (bottom), considering a frequency range between 116 and 132 MHz.} 
    \label{fig:med_502}
    \raggedright
    %\figurenote{\textit{Note.} In-situ measurements with high average (top), medium average (middle) and low average (bottom), considering a frequency range between 116 and 132 MHz.}
\end{figure}

To explore real-time data processing, we implemented a Savitzky--Golay filter with three levels of averaging. Figure~\ref{fig:med_502} shows filtered spectra with low (9-point polynomial), medium (6-point), and high (3-point) smoothing. Although computationally intensive, such averaging is useful for reducing data volume and identifying persistent spectral features in near-real time.\\

Each full-band acquisition, including averaged and unaveraged spectra and raw data files, required approximately 52.6~MB of storage. This metric informs future hardware and data management planning, especially as the system evolves toward longer integration times and field campaigns.\\

As a supplementary outcome of the project, six bilingual technical manuals were produced. These include guides for VNA calibration, dipole simulation in HFSS, and blade antenna modeling. The documentation is stored on the internal servers of the Universidad Industrial de Santander Research Groups E3T and RadioGIS for institutional access and training.

\section{Discussion}

\subsection{Implications for Sky Measurements}

The system temperature values in Table \ref{tab:temps} set practical limits on sensitivity for drift-scan observations. For integration times of several hours and moderate sky averaging, a noise floor of a few kelvin RMS per MHz bin is expected. While this is far from the 10--100~mK target required for direct detection of the global 21-cm signal, it is sufficient for characterizing foreground structure, verifying instrument stability, and developing calibration techniques.\\

\begin{table}[h]
\centering
\caption{Summary of system temperature and noise performance for three receiver configurations.}
\label{tab:temps}
\begin{tabular}{lccc}
\hline

\textbf{Configuration} & $T_\mathrm{SYS}$ (K) & $NF$ (dB) & $F$ (linear) \\
\hline

No preamplifier          & 1670 & 8.9 & 7.79 \\

Current preamplifier     & 453  & 4.1 & 2.56 \\

Filtered preamplifier    & 443  & 4.0 & 2.53 \\
\hline
\end{tabular}

\end{table}

Further reduction of system temperature would require either cryogenic front-end amplification or the use of differential radiometric techniques. These system temperature values (see Table \ref{table:11alturasbalun}) set practical limits on sensitivity for drift-scan observations. For the integration times estimated below, a noise floor of a few kelvin RMS per MHz bin is expected. While this is far from the 10–100 mK target required for direct detection of the global 21-cm signal, it is sufficient for characterizing foreground structure, verifying instrument stability, and developing calibration techniques before the next stage of observations with optimized antennas.

\subsection{Radiometric Sensitivity and Detection Thresholds}

To assess the theoretical sensitivity of the system in the context of global 21-cm signal detection, we apply the radiometer equation:

\begin{equation}
\Delta T = \frac{T_\mathrm{sys}}{\sqrt{B\,t}},
\end{equation}

where $\Delta T$ is the RMS thermal noise, $T_\mathrm{sys}$ is the system temperature, $B$ is the bandwidth, and $t$ is the integration time. Using the best-performing configuration described in Section~6, we adopt a measured system temperature of $T_\mathrm{sys} \approx 443$~K, representative of the filtered dual-stage preamplifier design.\\

Assuming a bandwidth of 1~MHz—sufficient to resolve the broad absorption feature predicted for the global 21-cm signal—and a total integration time of 1000 hours, the thermal noise limit is:

\begin{equation}
\Delta T = \frac{443}{\sqrt{10^6 \times 3.6 \times 10^6}} \approx 0.23~\mathrm{mK}.
\end{equation}

This result indicates that, under ideal observing conditions and assuming perfect calibration and system stability, the instrument is thermally capable of detecting signals at the sub-millikelvin level. Since the anticipated global 21-cm signal during the Cosmic Dawn has an amplitude on the order of 100~mK, the current system—at least from a thermal noise perspective—is sufficiently sensitive. For comparison, the HERA experiment used a similar integration time of 94 nights to place stringent upper limits on the 21~cm power spectrum during reionization, reaching sensitivities sufficient to rule out a broad class of cold IGM models \citep{thou}.\\

However, practical detection of the 21-cm signal requires suppression of instrumental systematics and astrophysical foregrounds to below the $\sim$10~mK level. In this context, beam chromaticity, gain instability, and calibration drift represent dominant challenges. Achieving effective signal recovery will require robust absolute calibration schemes, improved spectral smoothness, and potentially differencing or switching techniques to isolate sky signal from instrumental response.\\

The present sensitivity benchmark demonstrates that the system is capable of contributing to foreground characterization and site stability studies, and may serve as a viable platform for future calibrated global signal experiments. However, a useful future step in characterizing the system's long-term performance will be to conduct an Allan variance analysis of the radiometric output. This test provides a direct measure of the stability and noise behavior of the full analog and digital signal chain over varying integration times. This will give a measure of the timescale beyond which systematics such as temperature drift, gain fluctuations, or ADC clock instabilities begin to dominate over thermal noise, thus limiting the effective integration time. Performing this test in controlled laboratory conditions will help determine whether the final prototype can achieve the $\sim1000$-hour integrations required for global 21-cm signal detection, and will inform both hardware improvements and calibration strategies prior to field deployment.

\subsection{Comparison Between Simulations vs. Measurements}

\begin{figure}[h]
    \centering
    %\caption{$S_{11}$ simulation vs measurement plots (antena 1:1)} 
    \includegraphics[scale=0.45]{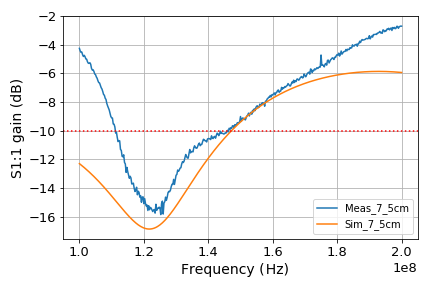}
    \includegraphics[scale=0.45]{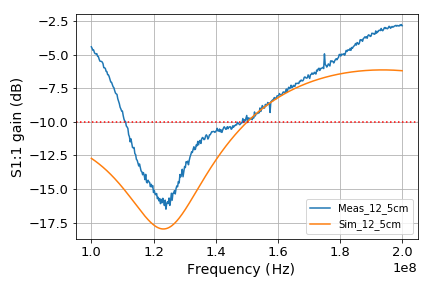}
    \includegraphics[scale=0.45]{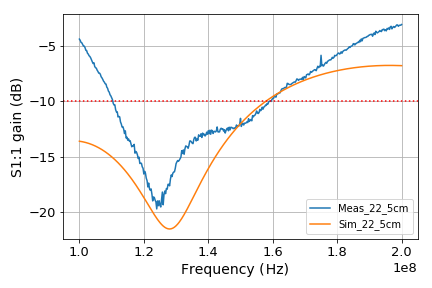}
    \includegraphics[scale=0.45]{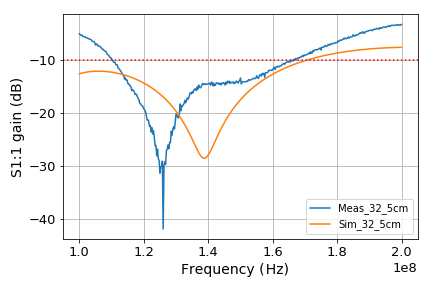}
    \includegraphics[scale=0.45]{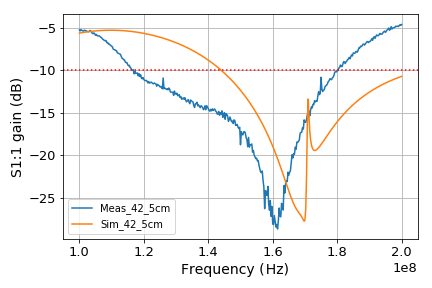} 

    \caption{$S_{11}$ simulation vs measurement plots for the blade antenna with balun tuner height @ 7.5 cm, 12.5 cm, 22.5 cm, 32.5 cm, and 42.5 cm.} 
    \label{fig:med_303}
    \raggedright
    %\figurenote{\textit{Note.} Simulation vs measurement comparison (top to bottom) @ 7.5 cm, 12.5 cm, 22.5 cm, and 32.5 cm balun tuner height. Ansys HFSS license from Universidad de Chile.}
\end{figure}

Here we address the consistency between simulation and measurement in terms of impedance match and gain, and assess receiver performance in terms of sensitivity, noise floor, and linearity. To do this we compare the $S_{11}$ response of the blade antenna across five balun tuner heights (VNA temperature stability: 0.05 dB/°C). Table~\ref{table:11alturasbalun} summarizes the key quantities from simulation and measurement, including the bandwidth defined by the $-10$dB threshold, the frequency of minimum reflection, and the peak return loss. Table \ref{table:11alturasbalun} illustrates the numeric values of this comparison.  The full comparison is shown in Figure \ref{fig:med_303}.\\

At low balun heights (7.5 and 12.5cm), simulations and measurements agree well, with the reflection minimum shifting by only 4 and 1MHz, respectively. As the balun height increases, the discrepancy becomes more noticeable. At 22.5, 32.5, and 45.5cm, the measured $S_{11}$ minima occur at lower frequencies than in simulation by 6, 13, and 8MHz, respectively. \\

The comparison between simulated and measured $S_{11}$ parameters shows in both the frequency of minimum reflection and the bandwidth. The modest but consistent deviations between simulations and measurements are likely due to limitations in the simulation setup. The HFSS model used a reduced ground plane (1.3x1.0 m aluminum sheet) to limit computational load, while the measured system includes a 5×5m wire mesh extension, which affects impedance matching—particularly at the lower end of the frequency band. \\

In addition, the simulations assume ideal material properties and perfectly symmetrical structures, whereas the real antenna may exhibit small asymmetries due to construction tolerances, feedline routing, and connector parasitics. Environmental factors such as ground conductivity, dielectric constant of the soil, and nearby objects may also influence the measured response. These effects result in frequency shifts of a few MHz in the location of the reflection minima, which are acceptable at this prototyping stage but will be addressed in future simulation campaigns incorporating full site-specific boundary conditions.

\begin{table}[h]
\centering
\caption{$S_{11}$ simulation vs measurement comparison for different balun heights.}
\label{table:11alturasbalun}
\begin{tabular}{lccccc}
\hline
\multicolumn{1}{c}{\textbf{Concept}}    & \textbf{7.5cm} & \textbf{12.5cm} & \textbf{22.5cm} & \textbf{32.5cm} & \textbf{45.5cm} \\ \hline

Simulation min. freq. @-10dB (MHz)      & 100            & 100             & 100             & 100             & 145             \\
 
Measurement min. freq. @-10dB (MHz)     & 112            & 112             & 111             & 111             & 117             \\

Simulation max. freq. @-10dB (MHz)      & 147            & 150             & 157             & 170             & 200             \\
 
Measurement max. freq. @-10dB (MHz)     & 145            & 149             & 158             & 166             & 179             \\

Simulation BW (bandwidth) (MHz)         & 47             & 50              & 57              & 70              & 55              \\
 
Measurement BW (bandwidth) (MHz)        & 33             & 37              & 47              & 55              & 62              \\

Simulation min. gain (dB)               & -16.58   & -17.98          & -21.25          & -28.54          & -27.24          \\
 
Measurement min. gain (dB)              & -15.81         & -16.50          & -19.72          & -41.91          & -28.45          \\

Simulation frequency @ min. gain (MHz)  & 122            & 123             & 130             & 139             & 169             \\
 
Measurement frequency @ min. gain (MHz) & 126            & 124             & 124             & 126             & 161            
\end{tabular} 
\\
\raggedright
\end{table}

\section{Future Work}

The results presented in this work establish the foundation for a low-frequency radiometric system capable of supporting future observational efforts targeting the global 21-cm signal and foreground removal strategies. Several lines of work are currently underway to improve the system's performance, expand its capabilities, and move toward field deployment in scientifically viable sites.\\

We are developing novel, software optimized antenna designs \citep{restrepo} along with radio sky model validation of antenna chromaticity and bayesian foreground models (Mora et al, in prep).\\

Another immediate objective is to improve the absolute calibration accuracy of the system. This includes implementing a three-position switching scheme (antenna, load, and noise diode) to enable calibration of the total power spectrum in physical temperature units, as well as tracking and correcting for gain fluctuations over time. Laboratory tests of the switching mechanism and noise injection levels are currently in preparation.\\

We are developing improved analog front-end systems that incorporate high-selectivity bandpass filtering and enhanced gain flatness across the 100--200~MHz band. This design will reduce out-of-band signal contamination and better match the dynamic range constraints of the SDR receivers. New amplifiers with lower intrinsic noise temperatures and better impedance matching are also being evaluated.\\

We are currently characterizing the RFI environment and antenna performance at potential observation sites, with short-term deployments at remote locations in the Colombian Andes and in Antarctica. Future deployments will focus on long-duration sky monitoring, interference analysis, and assessment of diurnal stability and ionospheric variability.\\

From a software perspective, future development will include real-time data compression and RFI excision algorithms implemented on embedded systems, in order to minimize storage requirements and support autonomous long-term operations. Real-time spectral averaging strategies, such as adaptive Savitzky–Golay filtering and time–frequency occupancy tracking, will be benchmarked against standard post-processing pipelines.\\

In the longer term, this platform is intended to serve as the core for a field-deployable, calibrated radiometer capable of foreground characterization and model testing for global 21-cm signal detection. In particular, it will contribute to the design and validation of instruments suitable for operation in extreme environments, including high-altitude sites in Colombia and future campaigns in the Antarctic Peninsula.\\

\section{Conclusions}

We have presented the design, simulation, implementation, and initial testing of a low-frequency radio telescope prototype developed to support future observations of the global 21-cm signal. The system combines a blade dipole antenna optimized for 100--200~MHz operation with a reconfigurable software-defined radio (SDR) receiver chain. Through a series of electromagnetic simulations, laboratory measurements, and in-situ tests, we evaluated the spectral response, sensitivity, and stability of the prototype in configurations relevant for global signal detection.\\

Simulations and measurements of the antenna reflection coefficient $S_{11}$ across multiple balun tuner heights demonstrated good agreement, with frequency shifts between modeled and measured resonances remaining below 15~MHz. SDR receiver tests showed that, while all devices offered basic functionality across the desired band, the Ettus E310 provided the best performance in terms of noise floor, linearity, and dynamic range. Cascaded low-noise amplifiers reduced the system temperature to approximately 450~K, confirming the viability of this analog front-end configuration for moderate-sensitivity drift scan experiments.\\

The integrated system was deployed for initial data acquisition in a rural site in the Eastern Colombian Andes. Spectral measurements across the 100--212~MHz range confirmed broadband functionality and supported the preliminary evaluation of analog and digital filtering strategies. Although the system is not yet optimized for the detection of the 21-cm cosmological signal, its current configuration enables site testing, calibration experiments, and foreground characterization.\\

Compared to established radiometers such as EDGES, SARAS, PRIZM, and BIGHORNS, this CANTAR prototype demonstrates broadly consistent performance in key areas such as impedance match and system temperature, while introducing a novel digital backend architecture based on commercial software-defined radios (SDRs). For instance, our best-performing antenna configurations achieve simulated and measured $S_{11}$ values below –25 dB over bandwidths of 50–70 MHz, which is comparable to the –15 to –20 dB targets reported for EDGES high-band \citep{bowman2009} and SARAS 3 \citep{Singh_2018}. Similarly, the prototype system temperature of 443–453 K using a dual-stage low-noise amplifier is within the 300–500 K range reported for BIGHORNS and PRIZM \citep{Sokolowski_2015, prizm}.\\

Our choice to use SDRs rather than custom analog signal chains or digitizers introduced higher noise floors and greater spectral instability in devices like the LimeNet Mini (noise floor ~–78 dB). This highlights the need to optimize gain settings to prevent compression or quantization artifacts. Clock jitter, ADC resolution limits, and undocumented internal filtering further complicated the calibration process—issues less prominent in fixed-hardware systems.\\

Despite these limitations, our results show that with proper analog filtering and preamplifier design, SDR-based systems can meet the core sensitivity and linearity requirements for global 21-cm observations. Radiometric sensitivity estimates indicate that, under ideal conditions, the current system could reach sub-millikelvin thermal noise levels over 1~MHz bandwidths with 1000 hours of integration, suggesting that—despite remaining challenges in calibration and systematics control—it is thermally capable of detecting global 21-cm signals of cosmological origin.\\

This prototype marks an initial but necessary step toward the deployment of scientifically capable instrumentation for global 21-cm cosmology in Colombia and beyond. Continued development will focus on improving calibration, stability, and RFI mitigation, while supporting training and capacity-building for the national radio astronomy community.\\

\section{Acknowledgements}

GC received support from Proyecto CODI 2021-42970 Convocatoria Proyectos de Investigación Regionalización 2021, Vicerrectoría de Investigación, Universidad de Antioquia. OR thanks the Radio Astronomy Instrumentation Group of Universidad de Chile for generously sharing their ANSYS license for the electromagnetic simulations. The CASIRI Station and the Antarctic scientific expeditions were funded by Vicerrectoría de Investigación y Extensión VIE at Universidad Industrial de Santander and by MinCiencias with the project "Desarrollo de un arreglo interferométrico de Radio Telescopios para establecer una estación de Radio Astronomía de la UIS en el Páramo de Berlín (Santander)". contract No. 82527 CT ICETEX 2022-0723. The authors also express their gratitude to Programa Antártico Colombiano, Comisión Colombiana del Océano, Fuerza Aeroespacial Colombiana, Armada de Colombia, Instituto Antártico Chileno, Ejército de Chile, Armada de Chile, Fuerza Aérea de Chile for all their support that made our Antarctic Expeditions possible. The authors thank Jimena Giraldo for typesetting assistance.

\bibliography{reference-file.bib}

\end{document}